\documentclass{aa}  
\usepackage{graphicx}
\usepackage{txfonts}
\usepackage[colorlinks]{hyperref}

\def\aap{A\&A}
\def\apj{ApJ}
\def\apjs{ApJS}

\def\hi{\ion{H}{i}}
\def\h2{H$_2$}

\def\kms{km\,s$^{-1}$}

\def\deg{\hbox{$^\circ$}}
\def\arcmin{\hbox{$^\prime$}}

\def\fdg{\hbox{$.\!\!^\circ$}}
\def\farcm{\hbox{$.\mkern-4mu^\prime$}}

\begin{document}

\title{\hi\ filaments are cold and associated with dark molecular gas}

   \subtitle{HI4PI-based estimates of the local diffuse CO--dark H$_{2}$
     distribution\thanks{FITS files for Figs. \ref{Fig_NH2_8} to \ref{Fig_NH_CO_90} are  available in electronic form
at the CDS via anonymous ftp to cdsarc.u-strasbg.fr (130.79.128.5)
or via {\url{http://cdsweb.u-strasbg.fr/cgi-bin/qcat?J/A+A/}} }}

   \author{P.\ M.\ W.\ Kalberla, \inst{1} J.\ Kerp, \inst{1} \and U.\ Haud
     \inst{2} }

\institute{Argelander-Institut f\"ur Astronomie,
           Auf dem H\"ugel 71, 53121 Bonn, Germany \\
           \email{pkalberla@astro.uni-bonn.de}
           \and
           Tartu Observatory, University of Tartu,
           61602 T\~oravere, Tartumaa, Estonia }

   \authorrunning{P.\,M.\,W. Kalberla, J.\ Kerp \& U.\ Haud } 

   \titlerunning{\hi\ filaments with diffuse H${_\mathrm{2}}$ }

   \date{Received 29 January 2020 / Accepted 30 April 2020 }

  \abstract 
{ There are significant amounts of \h2 in the Milky Way. Due to its
    symmetry \h2 does not radiate at radio frequencies. CO is thought to
    be a tracer for \h2; however, CO is formed at significantly higher
    opacities than \h2. Thus, toward high Galactic latitudes significant
    amounts of \h2 are hidden and are called CO--dark.} 
{We demonstrate that the dust-to-gas ratio is a tool for  identifying 
  locations and column densities of CO--dark \h2.} 
{We adopt the hypothesis of a constant $E(B-V)/N_{\mathrm{H}}$ ratio, 
  independent of phase transitions from \hi\ to \h2. We investigate the
  Doppler temperatures $T_{\mathrm{D}}$, from a Gaussian decomposition
  of HI4PI data, to study temperature dependences of
  $E(B-V)/N_{\mathrm{HI}}$. }
{The $E(B-V)/N_{\mathrm{HI}}$ ratio in the cold \hi\ gas phase is high
  in comparison to the warmer phase. We consider this as evidence that
  cold \hi\ gas toward high Galactic latitudes is associated with
  \h2. Beyond CO--bright regions,   for $T_{\mathrm{D}} \le
  1165~\mathrm{K}$ we find a correlation $(N_{\mathrm{HI}} +
  2N_{\mathrm{H2}})/N_{\mathrm{HI}} \propto -\log T_{\mathrm{D}}$. In
  combination with a factor $X_{\mathrm{CO}} = 4.0 \times
  10^{20}~\mathrm{cm}^{-2}~(\mathrm{K}~\mathrm{km\,s}^{-1})^{-1}$ 
  this yields  $N_{\mathrm{H}}/E(B-V) \sim
  5.1\ \rm{to}\ 6.7 \times 10^{21}~\mathrm{cm}^{-2}~ \mathrm{mag}^{-1}$ for the full sky,
  which is compatible with X-ray scattering and UV absorption line observations.}
{Cold \hi\ with $T_{\mathrm{D}} \le 1165~\mathrm{K}$ contains on average
  46\%  CO--dark \h2. Prominent filaments have $T_{\mathrm{D}} \le
    220~\rm{K}$ and typical excitation temperatures $T_{\mathrm{ex}} \sim
    50$ K. With a molecular gas fraction of $\ge61\%$ they are
    dominated dynamically by \h2.} 

  \maketitle
%
 
\section{Introduction}
\label{Intro}

The ISM is a multiphase medium and a major part consists of neutral
atomic and molecular gas that is highly intermixed with interstellar
dust \citep{Draine2003}. The most abundant atomic and molecular
constituents of the gas are \hi\ and \h2. \hi\ is easy to observe but
\h2 is homonuclear; it has no permanent electrical dipole moment, and
therefore rotational or vibrational transitions are not observable at
radio frequencies \citep{Carruthers1970}. Space-based far-UV
spectrographs are needed for \h2 observations \citep{Spitzer1959}. These
pencil-beam observations are involved, and therefore  supplementing
data from secondary tracers is often used to deduce the spatial and
density distribution of the \h2.  Because of its rather high abundance
and low excitation temperature, CO is considered to be a standard tracer
for molecular hydrogen. \h2 interacts with CO via collisions
\citep{Bolatto2013}, and therefore the CO line intensities and shapes are
quantitative measures for the \h2 volume density and kinetic gas
temperature. However, CO is formed at higher opacities and significantly
lower gaseous temperatures than \h2 \citep[][their
  Fig.\,1]{Bolatto2013}. Consequently, a major fraction of \h2 in the
local ISM is not associated with CO. We call molecular hydrogen CO--dark
when the CO does not trace it at all or the actual \h2 content exceeds
the amount expected from the observed CO and the standard
$X_\mathrm{CO}$ factor \citep{Bolatto2013}.

The only direct observational probes of \h2 in the diffuse ISM are the
far-UV electronic transitions in the Lyman and Werner bands
\citep{Spitzer1959}. To observe these lines space-based spectrographs
are needed, like that  on board   the {\it Copernicus} orbital observatory
or the {\it Far Ultraviolet Spectroscopic Explorer} (FUSE).  Pioneering
work was done by \citet{Savage1977} and \citet{Bohlin1978}, but only data
at a few hundred positions are available and the molecular gas fractions
$f^N_{H2} = 2 ~ N_{\mathrm {H2}} / (N_{\mathrm {HI}} + N_{\mathrm {H2}})
= 2 ~ N_{\mathrm {H2}} / N_{\mathrm {H}}$ for the diffuse medium in the
range $ 2~ 10^{19} < N_{\mathrm {H}} < 2~10^{21}~ {\rm cm^{-2}}$ are
particularly uncertain; we refer to Fig. 1 and further discussions of the
review by \citet{Snow2006}. These observation of \h2 toward
high Galactic latitudes indicate that major amounts of the molecular gas
are CO--dark. In addition,  cross-correlation
studies between different tracers of the total gas and the \hi\ column
density imply the existence of CO–dark \h2 (e.g., \citet{Reach1998},
  \citet{Planck2011}, \citet{Strong1996}, and \citet{Grenier2005}).

To estimate the amount of \h2 located away from CO--bright or even
star-forming regions, the tight correlation between the dust and gas is
of key interest. The dust far-infrared radiation
($I_\mathrm{FIR}$) and its optical extinction ($E(B-V)$) are both closely
correlated with $N_\mathrm{H}$ and must scale linearly with it
\citet{Liszt2014a,Liszt2014b}.  Using the linear correlation between 
$N_\mathrm{HI}$ and the optical extinction $E(B-V)$ \citep{Schlegel1998}
toward the low extinction regions of the high Galactic latitude sky,
\citet{Lenz2017} deduce $N_{\mathrm {HI}}/ E(B-V) = 8.8~ 10^{21}$
cm$^{-2}$~ mag$^{-1}$ at HI4PI angular resolution. With that value, they
derive a new map of interstellar reddening covering 39\% of the
sky. They need to restrict their approach to $N_{\mathrm {HI}} < 4~
10^{20} $ cm$^{-2}$ to prevent   opacity effects or phase transitions
from degrading the linear correlation.

Our aim here is to go beyond that limit in $N_\mathrm{HI}$. We do that
by accounting for the  CO--dark \h2. In the diffuse
ISM we show that a phase transition from \hi\ to \h2 does not have an immediate
feedback on the physical properties of the dust. We adopt the hypothesis
that the dust extinction  still scales linearly to $N_\mathrm{H}$
even when \h2 is forming in the diffuse ISM. When \h2 is formed the
\hi\ emission gets dimmer, but the optical extinction remains
unchanged. We use in the following $E(B-V)$ \citep{Schlegel1998} and
$N_\mathrm{HI}$ from HI4PI \citep{Winkel2016b} as observables, and when needed perform
  a consistency check with the $I_\mathrm{FIR}$ versus
$N_\mathrm{H}$ correlation. The difference between the dust traced
$N_\mathrm{H}$ and the observed $N_\mathrm{HI}$ is the (dark) amount of
$N_\mathrm{H_2}$. Because we restrict our investigation to high Galactic
latitudes ($|b| \ga 10\degr$), this minimizes the confusion with high
mass star-forming regions. Our approach is certainly not
straightforwardly applicable to these regions.

Toward the high Galactic latitude sky we find in the literature some spread
in ($N_{\mathrm {H}}/ E(B-V)$).  Using soft X-ray scattering
\citet{Predehl1995} find $N_{\mathrm {H}}/ E(B-V) = 5.55~ 10^{21}$
$\mathrm{cm}^{-2}$~mag$^{-1}$. At optical and UV wavelength
\citet{Savage1977} and \citet{Bohlin1978} determined $N_{\mathrm {H}}/
E(B-V) = 5.8~ 10^{21}$ $\mathrm {cm}^{-2}$~mag$^{-1}$ using Ly$\alpha$
and $N_{\mathrm {H2}}$ absorption against early-type stars. At radio wavelength \citet{Liszt2014a} deduces $N_{\mathrm {HI}}/ E(B-V)
= 8.3~ 10^{21}$ cm$^{-2}$~ mag$^{-1}$ from the LAB survey and
\citet{Lenz2017} find $N_{\mathrm {HI}}/ E(B-V) = 8.8~ 10^{21}$
cm$^{-2}$~ mag$^{-1}$ from HI4PI data.  These observations have differences of many
orders of magnitude  in wavelengths and also in the probed
spatial volumes and densities, but the $N_{\mathrm {HI}}/ E(B-V)$ ratio
is found to be remarkably constant. This implies that the gas-to-dust
ratio is not a function of the physical state of the gaseous phase
(e.g., gas temperature $T_\mathrm{gas}$, volume density $n_\mathrm{H}$, or
chemical composition $\Psi_\mathrm{H}$). These quantities
change with time, but    toward the high Galactic latitude
sky on the large angular scales probed by single dish \hi\ surveys, we know that the
physical conditions can be very closely approximated by a hydrostatic
equilibrium ansatz \citep{Kalberla2003}. The relevant parameter is only
the column density of hydrogen nuclei $N_\mathrm{H}$. This quantity
remains constant even during a phase transition from \hi\ to \h2.

Here we perform a cross-correlation analysis of neutral atomic gas (\hi)
and the interstellar reddening $E(B-V)$ toward the high Galactic
latitude sky. Toward these regions of interest we identify those
portions of the diffuse ISM which contain  CO--dark \h2. We
adopt the hypothesis that the dust-to-gas ratio is constant, or more
precisely that the optical extinction scales linearly with the column
density of the hydrogen nuclei. If this assumption is
valid, the Galactic foregrounds can be quantitatively evaluated
throughout the whole high Galactic latitude sky.

In Sect.\,\ref{Fitting} we investigate the correlation of the
interstellar reddening with the \hi\ gas temperature, extracted from a
Gaussian decomposition of the HI4PI survey. In Sect.\,\ref{Filaments} we
show a tight correlation of the \hi\ gas temperature along the major
axis of ISM filaments, which implies that the cold neutral medium (CNM) is
host to the  CO--dark \h2. Due to its low but
sufficient fraction of ionization \citep[][their Fig.\,1]{Crutcher2010}, the CNM is already 
closely interwoven with the magnetic lines of forces. In
Sect.\,\ref{Discussion} we put our findings in a perspective to the
debate on caustics in the ISM, and we focus on the question of whether the
$N_\mathrm{H}/E(B-V)$ ratio might depend on the gas temperature of the
CNM. We finish in Sect.\,\ref{Summary} with a brief summary
and some conclusions.

\section{$E(B-V)/N_{\mathrm  {HI}} $ dependences on \hi\ temperatures} 
\label{Fitting}

The temperature of a gaseous medium is characterized by thermal
motions. For \hi\ gas in equilibrium this kinetic temperature is related
to the spin temperature, which is the excitation temperature of the
hyperfine levels evaluated according to the Boltzmann equation
\citep{Field1959}. The 21 cm transition is usually collisionally
excited, and the spin temperature of the gas is a measure of the
kinetic temperature. To measure the spin temperatures it is necessary
to determine the 21 cm lines in both emission and absorption. This requires
sufficiently strong background sources and needs careful considerations
for radiative transfer effects along the line of sight
(e.g., \citet{Heiles2003a} and \citet{Murray2018a}). These investigations
are elaborate and are limited in practice to a small number of positions; for the millennium Arecibo 21 cm absorption-line survey, only
79 continuum sources were available \citep{Heiles2003a} and for the 21-SPONGE
\hi\ absorption line survey 57 lines of sight were available
\citep{Murray2018a}.

We use a Gaussian decomposition of the HI4PI survey and characterize the
temperature of an \hi\ cloud by its Doppler temperatures $T_{\mathrm
  {D}} = 21.86~ \delta v^2$ \citep[][Eq. 8]{Payne1980}. Here $\delta
v^2$ is the observed FWHM line width corrected for instrumental
broadening. Under typical conditions $T_{\mathrm {D}}$ is  a measure
  for an upper limit of the kinetic temperature $T_{\mathrm {kin}}$
(\citet{Field1959} and \citet{Field1969});   a coupling of the $\lambda$--21 cm excitation temperature
  and local gas motions is only possible   toward strong Ly--$\alpha$
  environments \citep{Liszt2001}.
  
 There is no unique use of the term Doppler temperature in the
 literature.  In the case of absorption lines \citet{Li2003} use the
 expression equivalent temperature. \citet{Heiles2003a} define a
 parameter $T_{\mathrm {k,max}}$ to describe the kinetic temperature
 of a component without nonthermal broadening  without naming
 $T_{\mathrm {k,max}}$ in a particular way. Nevertheless, this parameter
 is important for our understanding of the dynamical state of the
 ISM. Observed line widths result from intrinsic thermal broadening
 (representing kinetic temperatures $T_{\mathrm {kin}}$) and turbulent
 motions, causing the observed line broadening. This broadening,
 essentially resulting from a convolution of thermal and turbulent
 motions along the line of sight, is described by the characteristic
 turbulent Mach number \citep[][Sect. 6.2.4]{Heiles2003b}:
\begin{equation}
\label{eq:Mach}
M_{\rm t} = \sqrt{ 4.2 (T_{\rm D}/T_{\rm kin} - 1)}.
\end{equation}

Cold neutral medium clouds tend to be turbulent and supersonic, and high Mach numbers are
common \citep[][Fig. 12]{Heiles2003b}. In the CNM there is a
well-defined median magnetic field, and \citet{Heiles2005} conclude that
turbulence and magnetism are in approximate equipartition with a
characteristic turbulent Mach number $M_{\mathrm {t}} = 3.7$ at a median
CNM kinetic temperature of 50 K. Energy equipartition between magnetic
and kinetic energy implies $B^2/(4 \pi ) \propto n_{\mathrm {H}} \delta
v^2 $ for a hydrogen volume density $ n_{\mathrm {H}}$
(\citet{Crutcher1999}, \citet{Basu2000}, and \citet{Hennebelle2019}),
hence $B^2 \propto n_{\mathrm {H}} T_{\mathrm {D}}$.

The practical advantage of using Doppler temperatures is that
$T_{\mathrm {D}}$ can easily be determined from Gaussian components at
any observed position. This allows in particular a systematic
determination of Doppler temperatures along or across filaments that is
not possible with absorption data.  The CNM in filamentary structures
shows a well-defined log-normal distribution with a median $T_{\mathrm
  {D}} = 223$ K, corresponding to $M_{\mathrm {t}} = 3.7$ for the above-mentioned thermal temperature of 50 K
\citep[][Sect. 5.11]{Kalberla2018}. For a turbulent ISM that can be
described by a characteristic Mach number according to
Eq. \ref{eq:Mach}, $T_{\mathrm {D}}$ may therefore be considered as a
temperature measure. Uncertainties arise from unknown variations in
$M_{\mathrm {t}} $.

\subsection{Basic numerical strategies }
\label{Basics}

To study the dust-to-gas ratio we consider a constant ratio $ R =
E(B-V) / N_{\mathrm {H}} $, a cornerstone assumption according to
\citet{Liszt2014b}. Differently from all previous investigations, we
primarily do not study this ratio for integrated reddening or column
densities along the line of sight.  In general, $E(B-V)$ and
$N_{\mathrm {H}} $ may originate from several clouds or layers along the
line of sight.  We assume that \h2 forms out of the \hi\ phase and
variations of the dust-to-gas ratio are due to phase transitions from
one phase to the other. Decomposing the observed \hi\ line profile into
$n$ Gaussian components, we describe this configuration as
\begin{equation}  
E(B-V) = \sum_{i=1}^n  R ~ ( N_{\mathrm  {HI_i}} + 2~ N_{\mathrm  {H2_i}} )
= \sum_{i=1}^n  R ~ N_{\mathrm  {HI_i}}  ~ f_c(T_{\mathrm {D_i}})
\label{eq:EB}
,\end{equation}
and assign to each individual \hi\ cloud $i$ a conversion factor
$f_c(T_{\mathrm {D_i}})$ to take \h2 associated with this \hi\ cloud
into account.
The observed \hi\ may exist as a mixture of the cold, warm, and
lukewarm neutral medium (CNM, WNM, and LNM, respectively), in general with several such components along the
line of sight.  However it is expected that \h2 is associated only with
cold \hi\ (\citet{MO77}, \citet{Wolfire2003}, and \citet{Wolfire2010});
  this motivates us to assume that there is a temperature dependence of
  the correction factor $f_c(T_{\mathrm {D}})$.  In the simplest case,
  as assumed here, $f_c(T_{\mathrm {D}})$ may be independent of
  $E(B-V)$.

To determine $f_c(T_{\mathrm {D_i}})$ we need to develop a strategy for solving Eq. \ref{eq:EB}. We aim to tackle this iteratively by inserting
general accepted initial estimates from the literature (see 
next subsection), and then   improving the solution step by step.

Let us assume that we have some reasonable estimate
$f^{\mathrm{est}}_c(T_{\mathrm {D_i}})$. At a given position we separate
the Gaussian components (if present) in a selected range
$T_{\mathrm{D_{min}}} < T^{\mathrm{sel}}_{\mathrm {D}} < T_{\mathrm
  {D_{max}}}$. For \hi\ outside this $T^{\mathrm{sel}}_{\mathrm {D}}$
range we obtain the partial extinction as a sum of extinctions
for $m \le n$ components
\begin{equation}  
E^{\mathrm{part}}(B-V) = \sum_{i=1}^m  R ~ N_{\mathrm{HI_i}} ~ f^{\mathrm{est}}_c(T_{\mathrm {D_i}}).  
\label{eq:EBiter}
\end{equation}
For the selected Doppler temperature $ T^{\mathrm{sel}}_{\mathrm {D}} $
the observed column density is $N^{\mathrm{sel}}_{\mathrm {HI}} $ with a
corresponding extinction $(E(B-V) -
E^{\mathrm{part}}(B-V))$. The \hi\ based dust-to-gas ratio  is
\begin{equation}  
R_{\mathrm {HI}}(T^{\mathrm{sel}}_{\mathrm D}) = (E(B-V) - E^{\mathrm{part}}(B-V)) /
N^{\mathrm{sel}}_{\mathrm {HI}} .
\label{eq:RHI}
\end{equation}
When we have several components at similar Doppler temperatures
we treat them as a single component with $R_{\mathrm {HI}}(T^{\mathrm{sel}}_{\mathrm D})$. This
is equivalent to the assumption that these components have the same
ratio $R_{\mathrm {HI}}(T^{\mathrm{sel}}_{\mathrm D})$.

Relating $R_{\mathrm {HI}}(T^{\mathrm{sel}}_{\mathrm D})$ of the
selected component to the cornerstone ratio
\begin{equation}  
  R = (E(B-V) - E^{\mathrm{part}}(B-V)) / N^{\mathrm{sel}}_{\mathrm {H}},
\label{eq:N6}
\end{equation}
we have in agreement with Eq. \ref{eq:EB}
\begin{equation}  
R_{\mathrm {HI}}(T^{\mathrm{sel}}_{\mathrm D}) / R = N^{\mathrm{sel}}_{\mathrm  {H}} / N^{\mathrm{sel}}_{\mathrm
  {HI}} = (  N^{\mathrm{sel}}_{\mathrm  {HI}} + 2 N^{\mathrm{sel}}_{\mathrm {H2}} ) /  N^{\mathrm{sel}}_{\mathrm
  {HI}}  = f_c(T^{\mathrm{sel}}_{\mathrm {D}}). 
\label{eq:fCsel}
\end{equation}Thus, we use systematic deviations of the observed dust-to-gas ratio
$R_{\mathrm {HI}}$ from the cornerstone ratio $R$ to estimate the
associated \h2.  Applying this recipe to all observed positions in our
sample we derive distributions for $R_{\mathrm {HI}}(T_{\mathrm D})$ as
function of the selected Doppler temperatures $T^{\mathrm{sel}}_{\mathrm D}$.  In
principle the problem in deriving $f_c(T_{\mathrm {D}})$ is that we need
to start from scratch, $f^{\mathrm{est}}_c(T_{\mathrm {D_i}}) =
1$. Uncertainties in $f^{\mathrm{est}}_c(T_{\mathrm {D_i}})$ propagate
through Eq. \ref{eq:EBiter}.  We need to reiterate on
Eq. \ref{eq:EBiter}, successively improving
$f^{\mathrm{est}}_c(T_{\mathrm {D_i}})$. This task is considerably
simplified by taking some reasonable initial conditions into account.

\begin{figure}[th] 
   \centering
   \includegraphics[width=8.6cm]{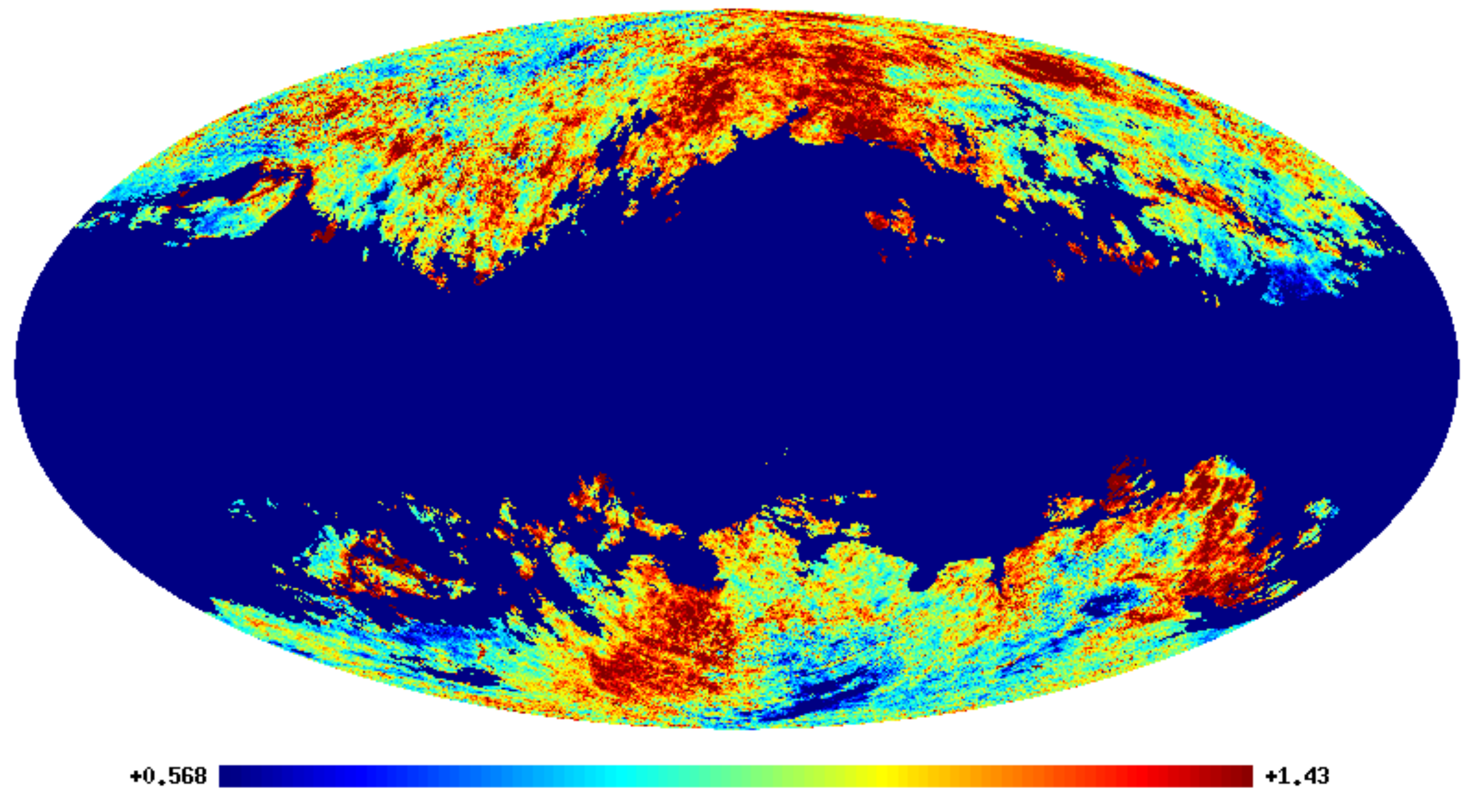}
   \includegraphics[width=8.6cm]{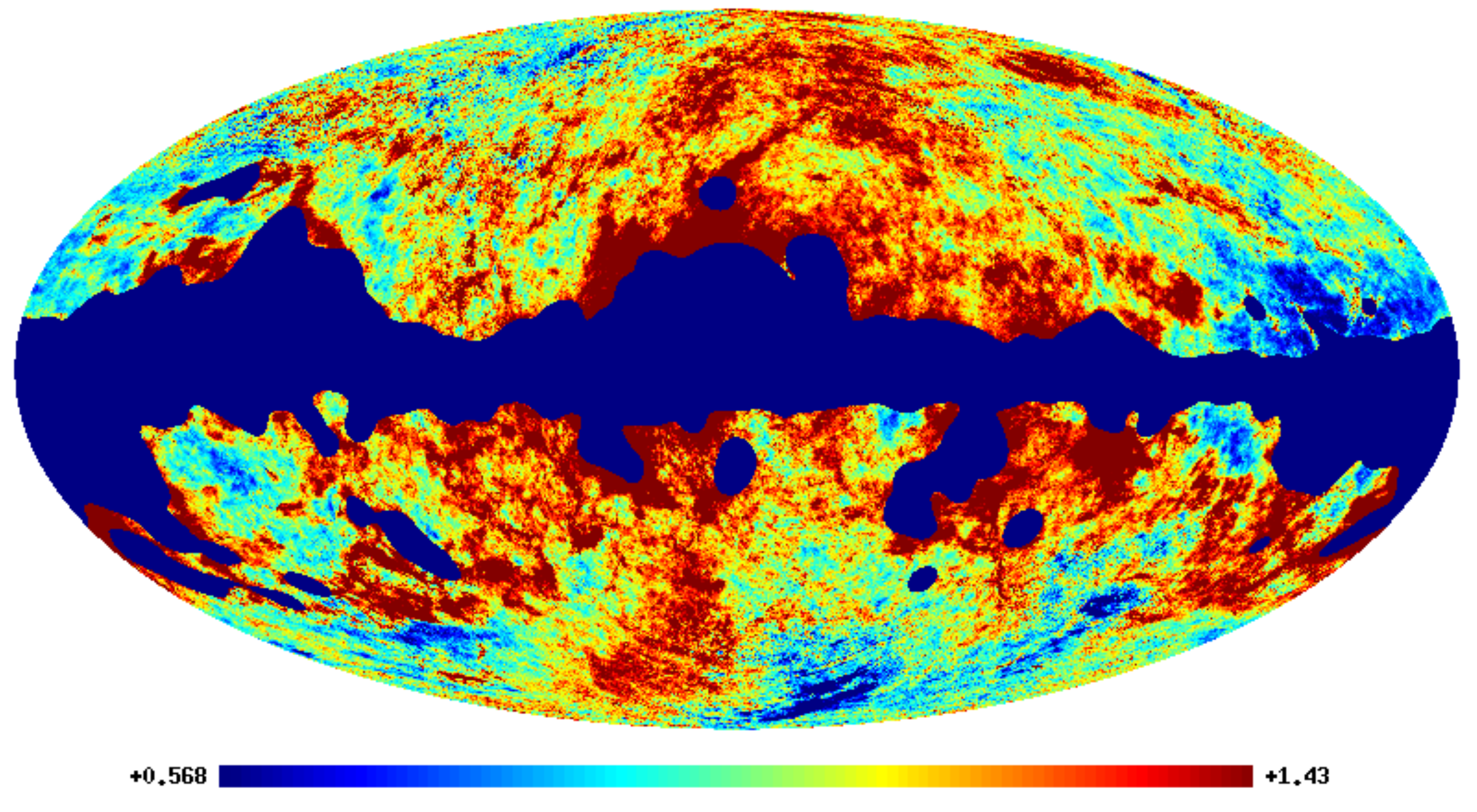}
   \caption{  $E(B-V)/N_{\mathrm {HI}}$ in regions used for the
     determination of the dust-to-gas ratio. Masked regions, including
     the Magellanic Clouds, are shown in  blue.  Top: $E(B-V)/N_{\mathrm
       {HI}}$ defined for canonical thin \hi\ gas with $N_{\mathrm {HI}}
     \le 4 ~ 10^{20}$ cm$^{-2}$ and $E(B-V) \le 0.08$ mag
     \citep{Lenz2017}. Bottom: $E(B-V)/N_{\mathrm {HI}}$ outside
     CO--bright regions. Units are $10^{-27}$ cm$^{2}$, scaling and
     color-coding   are as in Fig. \ref{Fig_Plot_EB_NH} for
     the mean corrected $E(B-V)/N_{\mathrm {H}}$ dust-to-gas ratio
     within a 2$\sigma$ range. } 
   \label{Fig_mask}
\end{figure}

\subsection{Bootstrap, initial constraints}
\label{Initial}

To simplify a solution of Eq. \ref{eq:EBiter} we make use of several
previously published results and adopt some broadly accepted
assumptions: 

\begin{enumerate}
\item $E(B-V)/N_{\mathrm {HI}}$ is well defined for $N_{\mathrm {HI}}
  \la 4 ~ 10^{20}$ cm$^{-2}$ and $E(B-V) \la 0.08$ mag
  (e.g.,  \citet{Savage1977}, \citet{Liszt2014a,Liszt2014b}, and
  \citet{Lenz2017}). We limit
  our first attempts to solve Eq. \ref{eq:EB} for this range 
  referred to in the following as canonical thin gas, see
  Fig. \ref{Fig_mask} top. This is the case considered by
  \citet[][Fig. 9]{Lenz2017}, and we use their ratio of
  $R = E(B-V)/N_{\mathrm {HI}} = (1.113 \pm 0.002) \times 10^{-22}$
  cm$^{2}$~mag as the cornerstone dust-to-gas ratio.
\item $E(B-V)/N_{\mathrm {HI}}$ is best defined toward high latitudes; a
  significant onset of \h2 formation is expected in the range $20\deg
  \ga |b| \ga 8\deg$ (\citet{Liszt2014a,Liszt2014b} and
  \citet{Dame2001}). We avoid $|b| \la 8\deg$ since there are no sight
  lines with small $E(B-V)$ in this range.
\item It appears well established that the molecular hydrogen fraction
  $f^N_{H2} = 2 ~ N_{\mathrm {H2}} / N_{\mathrm {H}} = 1 - 1/f_c
  (T_{\mathrm {D}})$ for the WNM should be negligible (\citet{MO77} and
  \citet{Wolfire2003}). Initially we define $ f_c(T_{\mathrm {D}}) = 1 $
  for the WNM.
\item The \h2 distribution may be CO--dark \citep{Grenier2005}. There is
  a fundamental difference between \h2 associated with CO in dense
  molecular clouds (CO--bright \h2) and \h2 outside such clouds where the
  gas-phase carbon resides in C or C$^+$ (CO--dark \h2)
  \citep{Wolfire2010}.  We exclude CO--bright regions with observed CO
  emission. To generate  CO--masked regions for exclusion, we identify the
  areas by using the type 2 product from the {\it Planck} legacy data
  release\footnote[1]{
    \url{https://wiki.cosmos.esa.int/planckpla2015/index.php/CMB_and_astrophysical_component_maps}}.
  We smooth these data heavily with a 5\deg ~Gaussian beam and mask all
  data for CO emission above a level of 0.2 K. Our CO mask covers 25.9\%
  of the sky, compared to \citet{Planck2014} who used, without smoothing,
  a sky fraction with significant CO emission in excess of 0.15 K \kms\
  (about 18\% of the sky). We mask additional regions that are affected
  by the Magellanic Clouds to avoid outliers in the  dust-to-gas ratio
  by contamination from these sources, see Fig. \ref{Fig_mask} bottom.
\item \hi\ gas at high velocities contains insignificant amounts of dust
  (\citet{Wakker1997} ,\citet{Miville2005}, and \citet{Lenz2017}).  The
  main body of the dust-bearing gas is associated with velocities
  $|v_{\mathrm {LSR}}|\la 90 $ \kms\ and we use this velocity range.
\item Single-dish \hi\ data may suffer from unknown optical depth
  effects. We consider corrections as proposed by \citet{Lee2015} and
  \citet{Murray2018} by multiplying observed column densities by
  a factor $f = \mathrm{log_{10}}( N_\mathrm{HIobs}/ 10^{20}) \times
  (0.26 \pm 0.02) + (0.91 \pm 0.02)$ or alternatively by $f =
  \mathrm{log_{10}}( N_\mathrm{HIobs}/ 10^{20}) \times (0.19 \pm 0.02) +
  (0.89 \pm 0.02)$ \citep{Nguyen2018}.
  \end{enumerate}
During iterations, after obtaining a reasonable accurate solution of
Eq. \ref{eq:EBiter} some of these estimates and constraints can be
released. We discuss this later in context.

\begin{figure}[thp] 
   \centering
   \includegraphics[width=8.7cm]{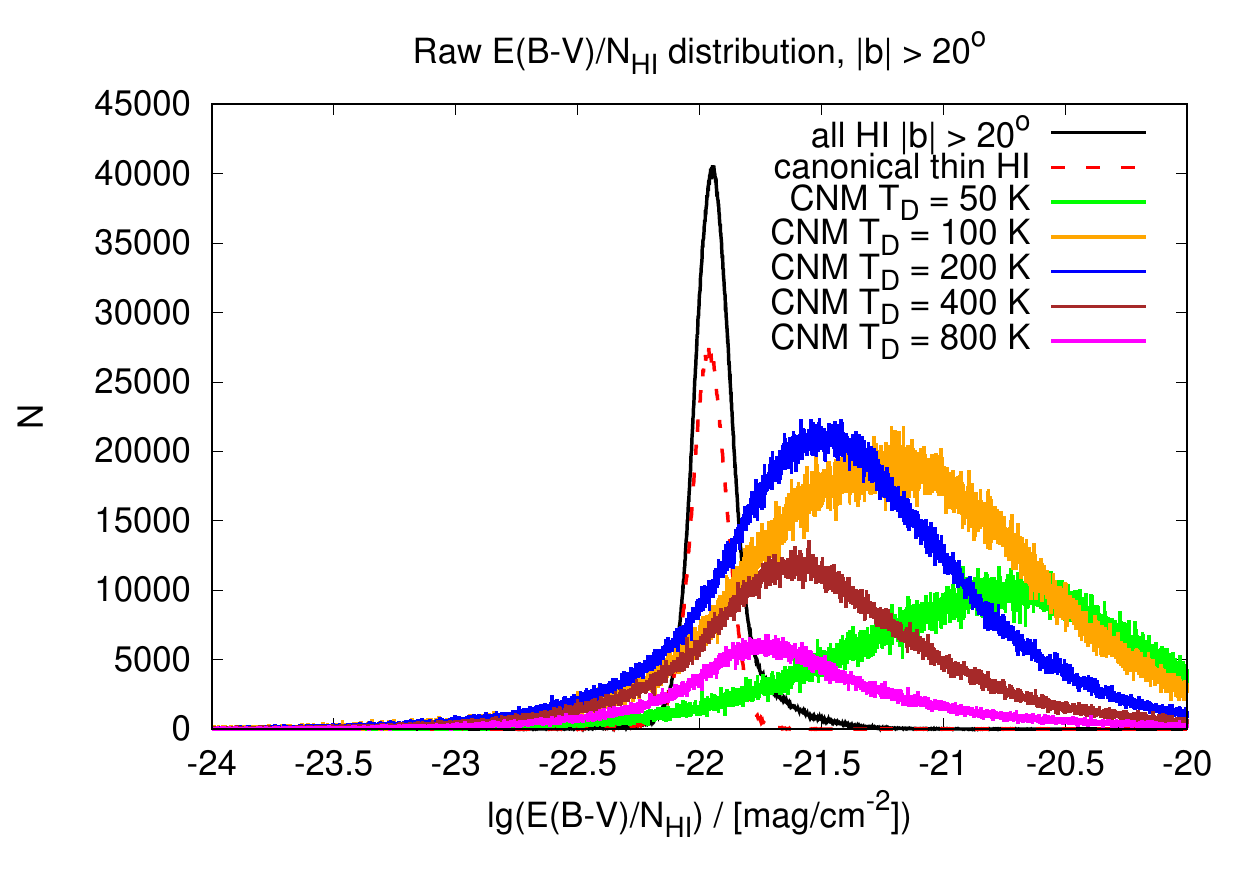}
   \includegraphics[width=8.7cm]{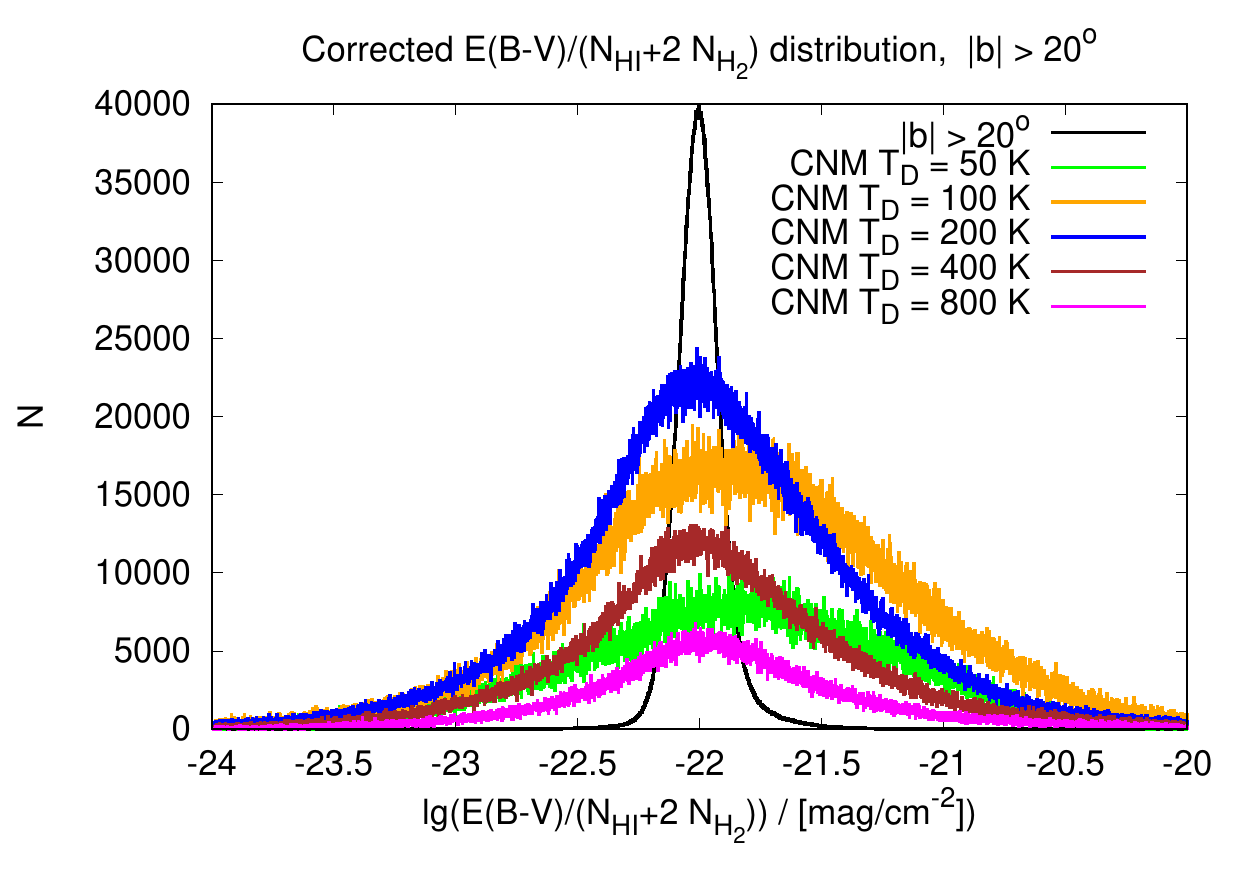}
   \includegraphics[width=8.7cm]{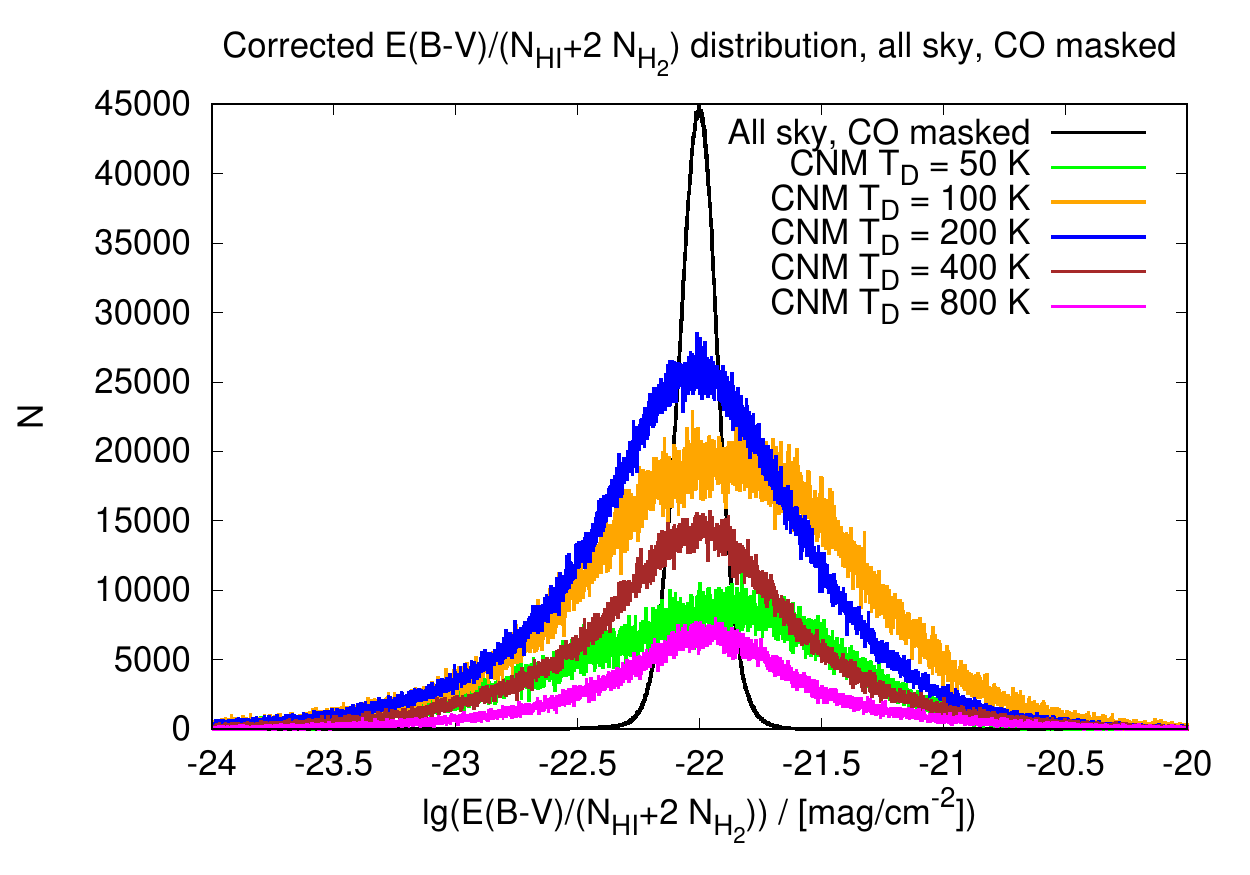}
   \caption{ Probability distribution functions for Gaussian
       components with selected Doppler temperatures
       $T^{\mathrm{sel}}_{\mathrm D}$, top: $E(B-V)/N_{\mathrm {HI}}$
       for latitudes $|b| > 20\deg$ (black)  and for canonical thin
     \hi\ (red dashed line) \citep{Lenz2017} compared to distributions of
     partial $E(B-V)/N_{\mathrm {HI}}$ ratios at Doppler temperatures
     $T_{\mathrm {D}} = $ 50, 100, 200, 400, and 800 K. Middle: $
     E(B-V)/( N_{\mathrm {HI}} + 2 N_{\mathrm {H2}})$ distributions for
     $|b| > 20\deg$. All components (black line) in comparison to scaled
     distribution at selected Doppler temperatures.  Bottom: $E(B-V)/(
     N_{\mathrm {HI}} + 2 N_{\mathrm {H2}})$ distributions now all-sky,
     but excluding CO--bright regions. The amplitudes of the partial
     distributions are scaled up by constant factors.  }
   \label{Fig_EB_NH}
\end{figure}

\subsection{Databases }
\label{Data}

We use HI4PI \hi\ data decomposed into Gaussian components as described
by \citet{Kalberla2018}. This decomposition is limited by the blending
of the obtained Gaussians, and in most cases the decomposition toward a
single line of sight is not unique (Sect. 4.1 of \citet{Haud2000}).
However, it is feasible to deduce a reliable Gaussian decomposition by
accounting not for isolated line profiles, but for larger neighboring
regions. This enables us to find a coherent solution for the whole area of
interest and yields more statistically independent Gaussian components
for corresponding lines of sight.

The largest groups of blended Gaussians in each profile from the
unmasked region of the sky in the bottom panel of Fig. \ref{Fig_mask}
contain on average 5.7 Gaussians. The uniqueness and stability of our
decompositions of the profiles, represented by six blended Gaussians, has
been examined in Sect. 4.2 of \citet{Haud2000}. The results of the
modeling of the decomposition process  demonstrate that the
blending is mostly a problem for broad HI lines, while the CNM with its
sharply defined narrow lines allows unique clues. As  we are mostly
discussing the cold gas here, we consider the used decomposition results
appropriate for the present statistical study.

The interstellar reddening $E(B-V)$ data are from
\citet{Schlegel1998}\footnote[2]{\url{https://lambda.gsfc.nasa.gov/product/foreground/fg_sfd_get.cfm}}.
We resample this data set to a homogeneous HEALPix grid
\citep{Gorski2005} with nside = 1024 and apply the correction $E(B -
V)_{\mathrm{true}} = 0.884 ~ E(B-V)_{\mathrm{downloaded}}$
\citep{Schlafly2011}, consistent with the scaling used by
\citet{Lenz2017}. Our analysis is in all cases done on a HEALPix grid
with nside = 1024. We use the tool ud\_grade from the HEALPix software
distribution\footnote[3]{\url{https://sourceforge.net/projects/healpix/}}
for up- or downgrading. The \hi\ data have a FWHM resolutions of
10\farcm8 for the northern sky \citep{Winkel2016} and 14\farcm5 for the
southern  \citep{Kalberla2019}. Column densities are accurate to
2.5\% \citep{Winkel2016}. The $E(B - V)$ data have a resolutions close
to 7\arcmin\  for a HEALPix grid with nside = 512 and are assumed to be
accurate to 16\% \citep{Schlegel1998}.

\subsection{Fitting $f_c(T_{\mathrm {D}})$}
\label{Fitting_fc}

We use a limited sample of selected Doppler temperatures
$T^{\mathrm{sel}}_{\mathrm {D}}$ to determine the dust-to-gas ratios
$R_{\mathrm {HI}}(T^{\mathrm{sel}}_{\mathrm D})$ according to Eq. \ref{eq:RHI} across
the sky. Our approach is iterative: initial estimates of
$f_c(T_{\mathrm {D}})$ were restricted to the canonical thin sample;
later we released constraints as far as possible. Here we demonstrate our
results for the final $f_c(T_{\mathrm {D}})$ fit from Eq. \ref{eq:f_c}.

The top panel of Fig. \ref{Fig_EB_NH} displays the log-normal
distributions for $R_{\mathrm {HI}}(T^{\mathrm{sel}}_{\mathrm D})$ for a
range of Doppler temperatures, $T^{\mathrm{sel}}_{\mathrm {D}} = $ 50,
100, 200, 400, and 800 K. For decreasing $T_{\mathrm {D}}$ the ratios
$R_{\mathrm {HI}}(T^{\mathrm{sel}}_{\mathrm D})$ shift to higher
values. In the top panel of Fig. \ref{Fig_EB_NH} we display for
comparison the ratios $E(B-V) / N_{\mathrm {HI}}$ for the total
reddening and integrated column densities for all data with $|b| >
20\deg$ (black) and the canonical sample according to \citet{Lenz2017}
(red dashed). Extending the sample to a larger fraction of the sky leads
to an extended asymmetric wing of $E(B-V) / N_{\mathrm {HI}}$ for $|b| >
20\deg$. Most of the \hi\ gas belongs to the WNM, the dust-to-gas
ratio for this part is unaffected. Deviations in $E(B-V) / N_{\mathrm
  {HI}}$ are caused by the CNM. The $T^{\mathrm{sel}}_{\mathrm
  {D}}$-selected $R_{\mathrm {HI}}(T^{\mathrm{sel}}_{\mathrm D})$
samples are scaled up in amplitude, but belong to the extended wing of
the black $E(B-V) / N_{\mathrm {HI}}$ distribution. The amplitudes of
the selected $R_{\mathrm {HI}}(T^{\mathrm{sel}}_{\mathrm D})$
distributions reflect the frequency distribution of components with
different Doppler temperatures or line widths
\citep[][Fig. 4]{Kalberla2018}. CNM components with $T_{\mathrm {D}}
\sim 220$ K are most frequent \citep[][Fig. 13]{Kalberla2016}.

\begin{figure}[thp] 
\centering \includegraphics[width=9cm]{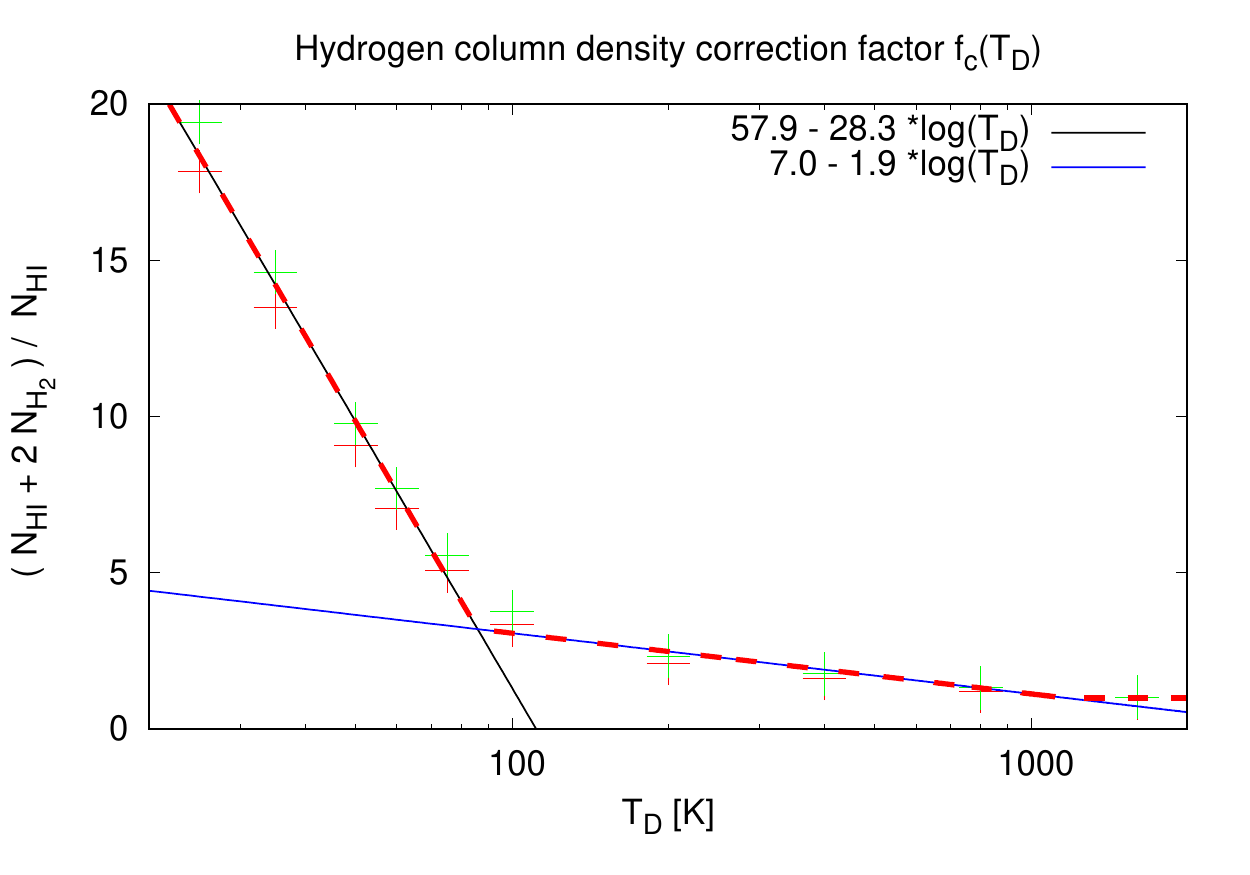}
   \caption{Hydrogen column density correction factor $f_c(T_{\mathrm
       {D}}) $ from an  all-sky fit, excluding positions in CO--bright
     regions. The crosses show fit results from the last 
     iterations, the solid lines the regressions from Eq. \ref{eq:f_c},
     the dashed red line the adopted complete solution. }
   \label{Fig_Tss_fit}
\end{figure}

We use the $R_{\mathrm {HI}}(T^{\mathrm{sel}}_{\mathrm D})$
distributions to determine $f_c(T_{\mathrm {D}})$ according to
Eq. \ref{eq:fCsel}; this is simply the factor needed to shift each
log-normal distribution in Fig. \ref{Fig_EB_NH}, top, to the canonical
$R$-value, the geometrical mean of the canonical $E(B-V) / N_{\mathrm
  {HI}}$ distribution (red). For each of the log-normal distributions we
determine a mean correction factor $\langle
f_c(T^{\mathrm{sel}}_{\mathrm {D}}) \rangle$
from the geometrical mean of the $ (E(B-V) - E^{\mathrm{part}}(B-V)) /
N^{\mathrm{sel}}_{\mathrm {HI}}$ distribution by fitting a
Gaussian. After a few iterations it became clear that this correction
can  be approximated astonishingly well by only two regression lines,
linear fits for $\log (T_{\mathrm {D}}):$
\begin{align}
 f_c(T_{\mathrm {D}}) & = 57.9 - 28.3 \times \log(T_D) & \text{
   for $T_{\mathrm {D}} < 85$ K} \nonumber \\
  & = 7.0 - 1.9 \times \log(T_D) & \text{ for $85 \le T_{\mathrm
     {D}} \le 1165 $ K} \nonumber \\
  & = 1.0 & \text{ for $T_{\mathrm {D}} > 1165 $ K}.
 \label{eq:f_c}
\end{align}
This best fit result to these two linear regressions is found by
selecting all positions outside CO--bright regions. The convergence of
the fitting process is slow, with oscillations around a single dominant
pole but decreasing amplitudes for the deviations between two successive
iterations. After seven unconstrained iterations we decided to terminate
this process by using the mean of two successive iterations as a
penalty. We display in Fig. \ref{Fig_Tss_fit} our final result. The
solid lines represent the two derived regressions; we also plot   the
bracketing results from the last two iterations (green and red
crosses). The dashed line gives the combined $f_c(T_{\mathrm
  {D}})$-solution according to Eq. \ref{eq:f_c} that we use in the
following.

\begin{figure*}[tbp] 
    \centering
    \includegraphics[width=9cm]{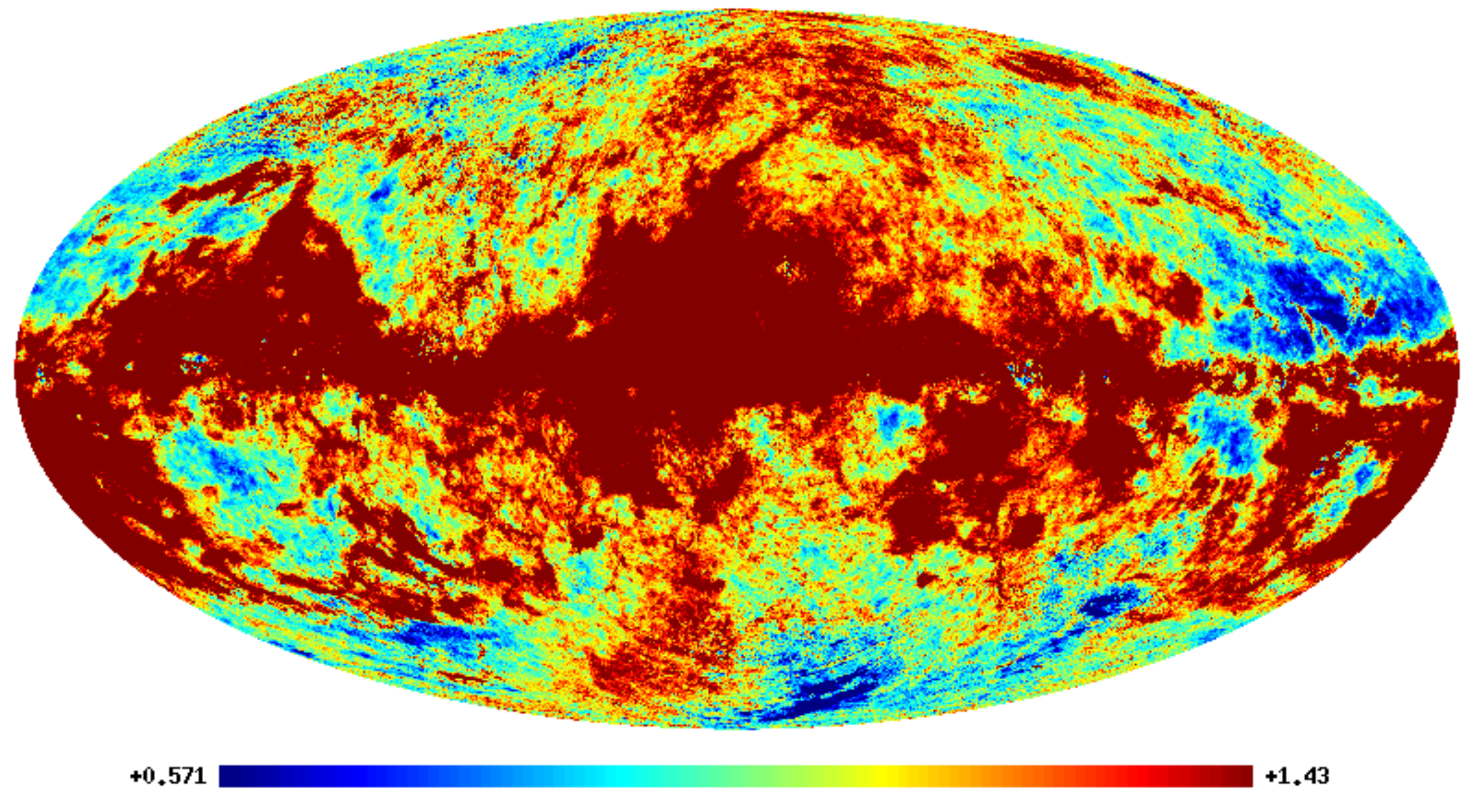}
    \includegraphics[width=9cm]{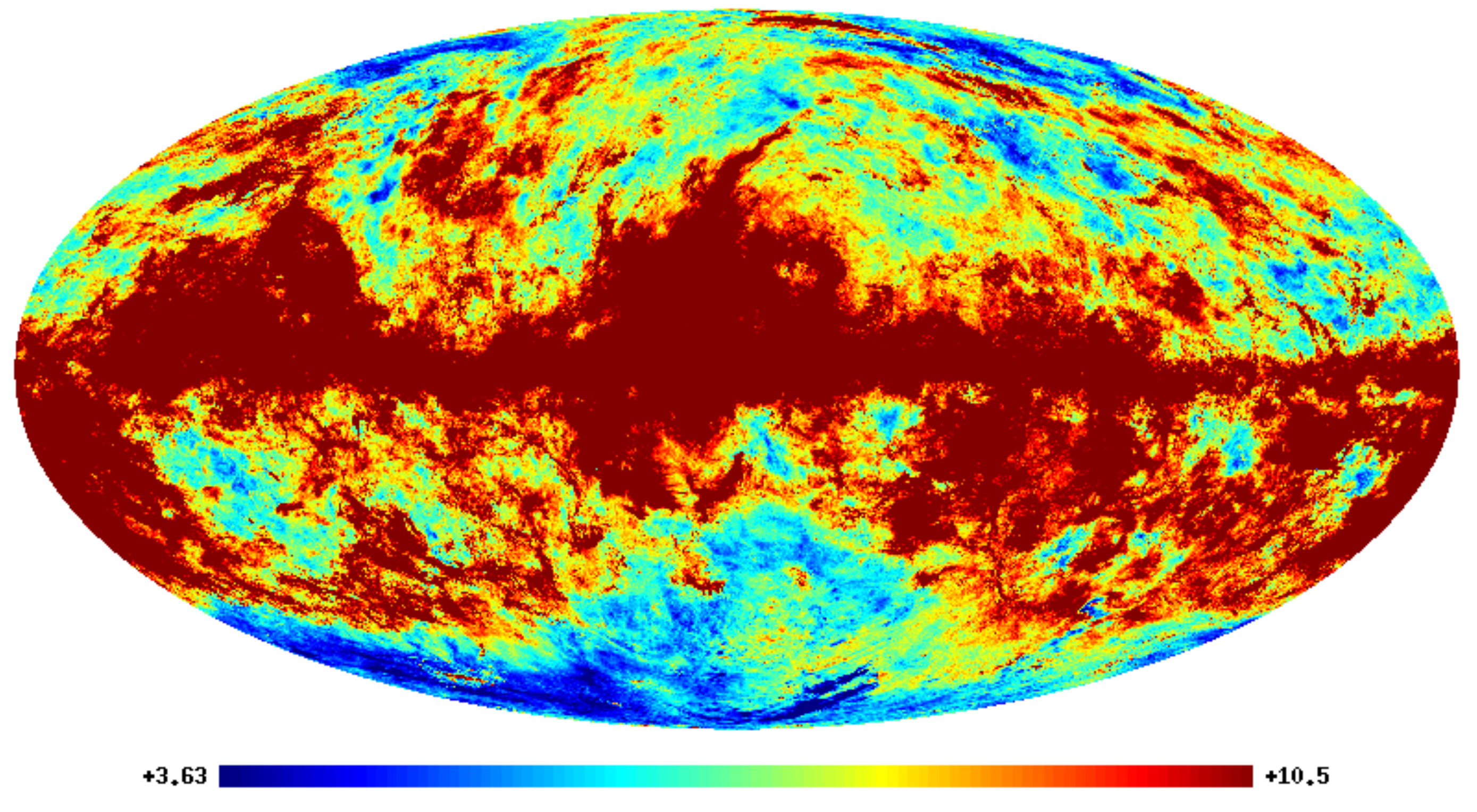}
    \includegraphics[width=9cm]{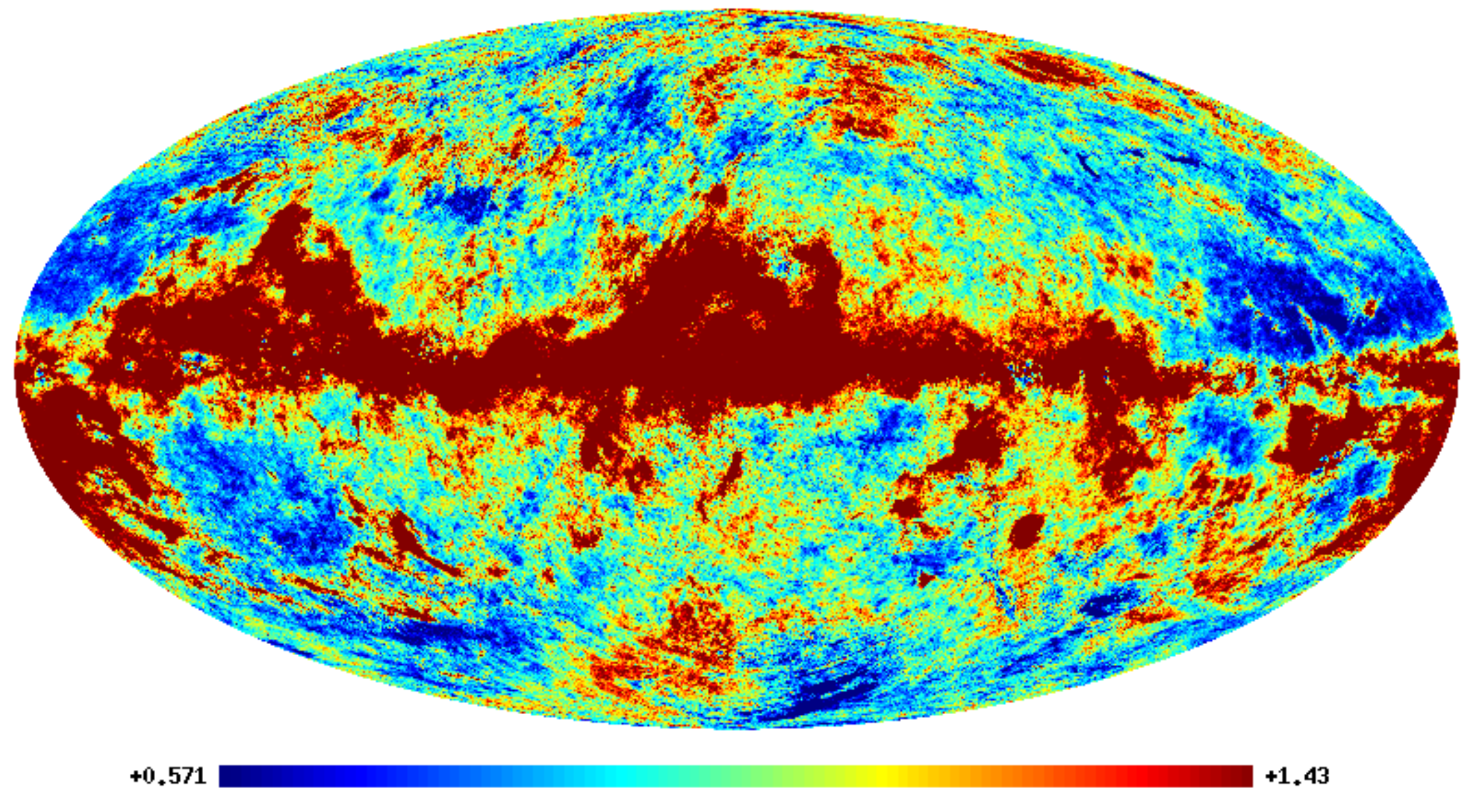}
    \includegraphics[width=9cm]{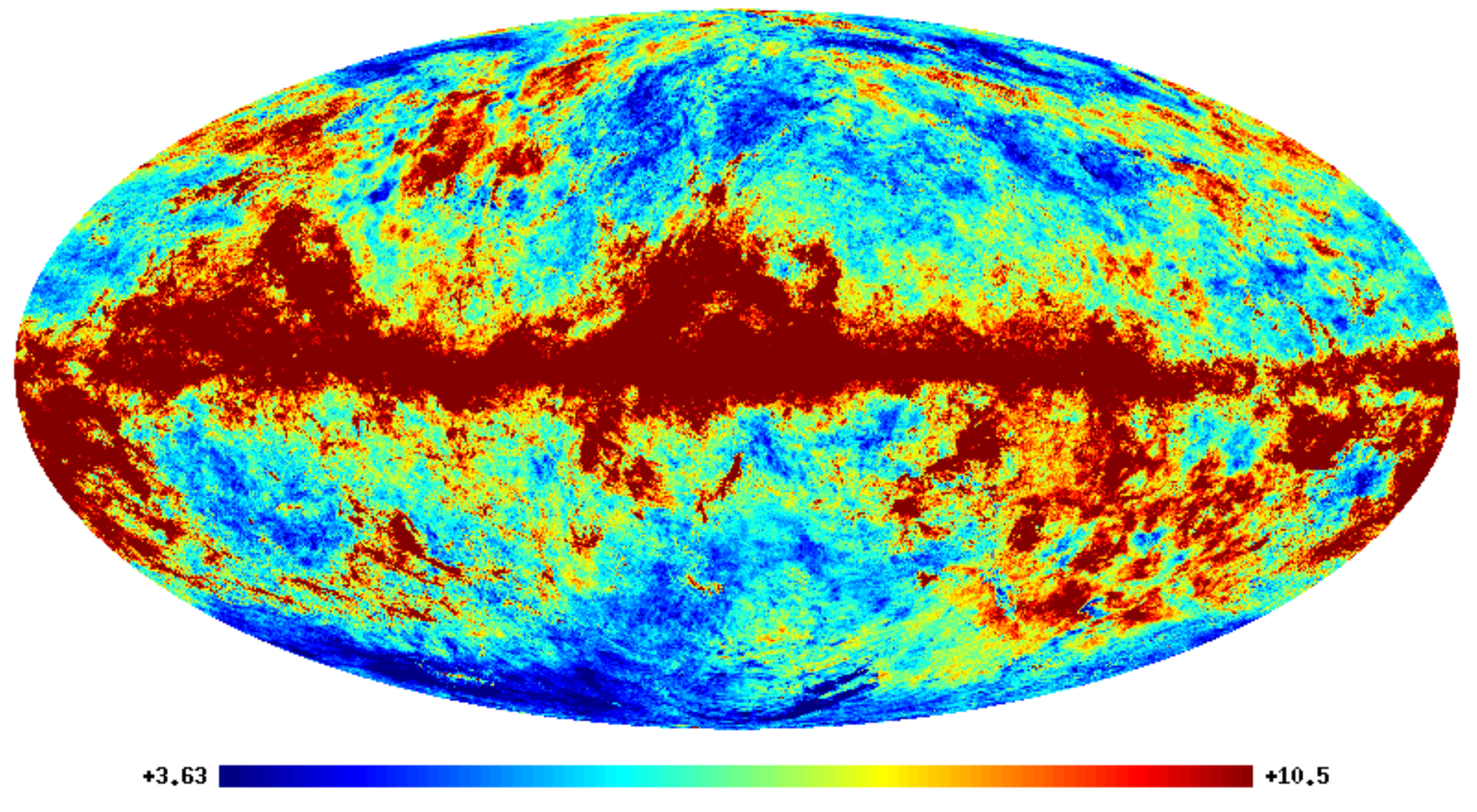}
    \includegraphics[width=9cm]{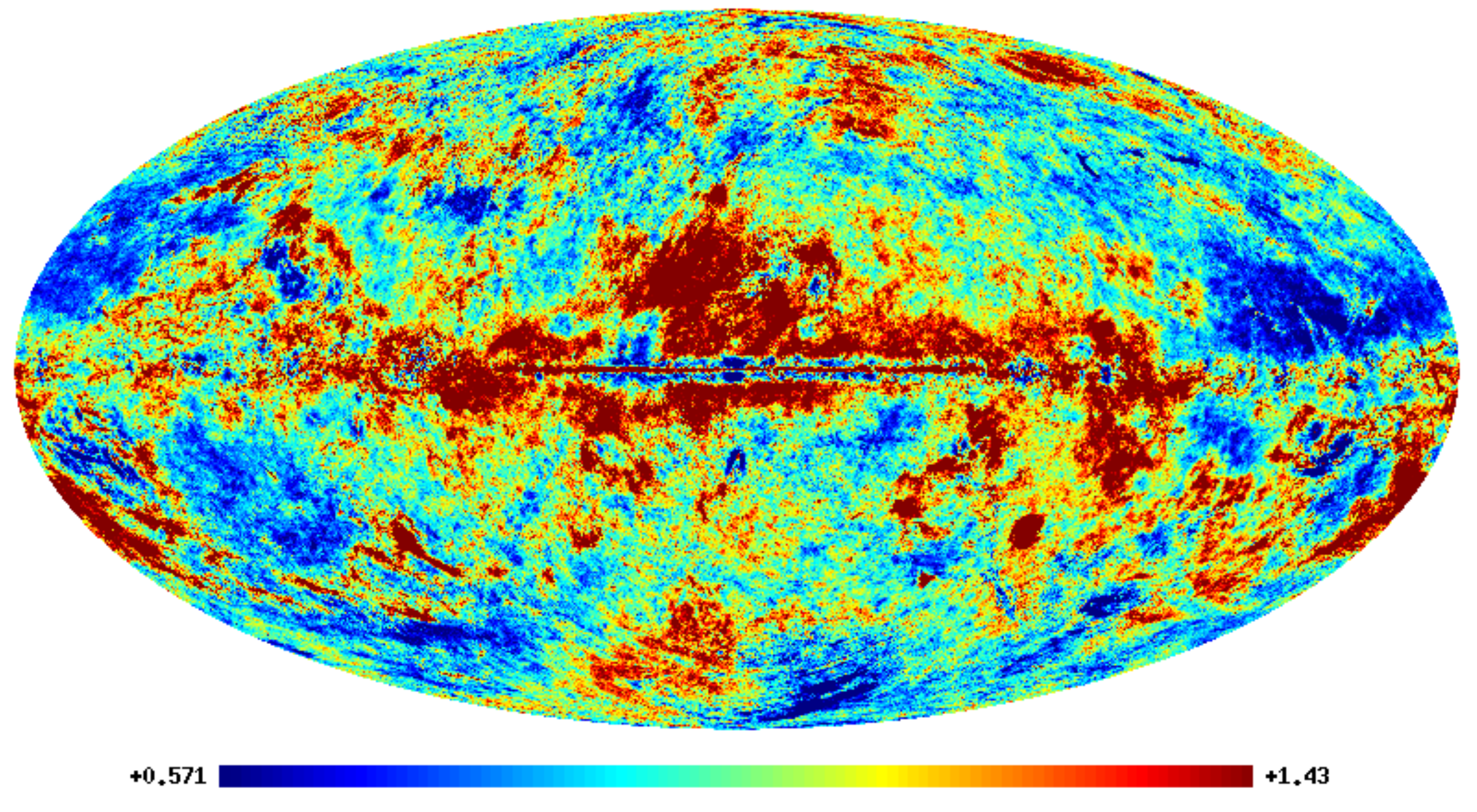}
    \includegraphics[width=9cm]{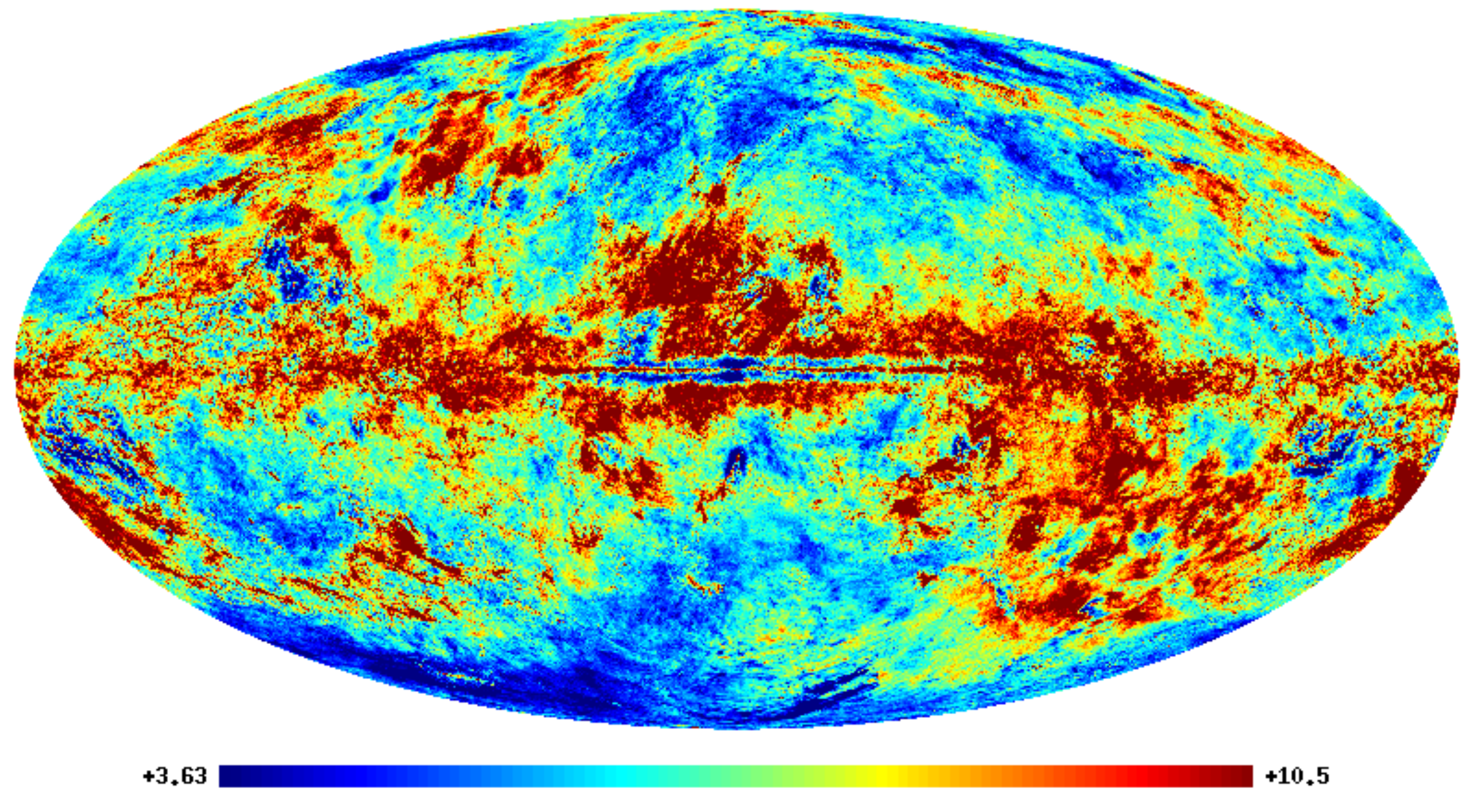}
    \caption{Left: $E(B-V)/N_{\mathrm {HI}}$ (top),
      $E(B-V)/N_{\mathrm {H}}$ with the $f_c(T_{\mathrm {D}})$
      correction EQ. \ref{eq:f_c} (middle), and $E(B-V)/N_{\mathrm
        {Htot}}$ including a correction for $X_{\mathrm {CO}} = 4.0
      \times 10^{20}$ cm$^{-2}$ (K \kms)$^{-1}$ (bottom). Units are
      $10^{-22}$ cm$^{2}$~mag, displayed are data around the mean within
      a 2$\sigma$ range on both sides. For a comparison of
      $E(B-V)/N_{\mathrm {HI}}$ after restricting the data to the
      canonical thin gas and to CO--bright regions we refer to
      Fig. \ref{Fig_mask}. Right: $\tau_{353}/N_{\mathrm {HI}}$
      (top), $\tau_{353}/N_{\mathrm {H}}$ (middle), and
      $\tau_{353}/N_{\mathrm {Htot}}$ (bottom).  Units are $10^{-27}$
      cm$^{2}$, also displayed are data within 2$\sigma$
      around the mean. }
   \label{Fig_Plot_EB_NH}
\end{figure*}

In the middle and lower panels of Fig. \ref{Fig_EB_NH} we display with
the same color-coding as on top the CNM distributions of $(E(B-V) -
E^{\mathrm{part}}(B-V)) / N^{\mathrm{sel}}_{\mathrm {H}}$ after applying
the $f_c(T_{\mathrm {D}})$-correction from Eq. \ref{eq:f_c}. In addition
we display the $E(B-V) / N_{\mathrm {H}}$ distributions (black)
calculated from total reddening and integrated hydrogen column densities
after application of the $f_c(T_{\mathrm {D}})$-correction. The results
in the middle panel are for $|b| > 20\deg$, at the bottom for all-sky
but excluding CO--bright regions (see Fig. \ref{Fig_mask}). For both
panels the CNM distributions are centered close to the canonical $R =
E(B-V)/N_{\mathrm {HI}} = 1.113 \times 10^{-22}$ cm$^{2}$~mag. The
$E(B-V) / N_{\mathrm {H}}$ distribution in the bottom panel has a nearly
Gaussian shape, without the extended wings   seen in the upper panels. This
is our best fit result; the range $|b| > 20\deg$ still contains 
 some contaminations from CO--bright regions. 

During the iteration procedure it became clear that
$f_c(T_{\mathrm{D}})$ can be consistently deduced by dropping the major
constraints considered in Sect. \ref{Initial}. Constraints on column
densities or interstellar extinction (items 1 and 2) are not necessary. It is very
important,  however, to distinguish between CO--dark and CO--bright
regions (item 4); it even appears necessary to extend the mask for the
CO--bright parts of the sky spatially by smoothing the observed CO
emission heavily. The selection of the velocity range (item 5)   has
a very limited impact on the fit results.  It is not necessary to
distinguish \hi\ phases (item 3); the $f_c(T_{\mathrm{D}})$ solution
according to Eq. \ref{eq:f_c} applies only to $T_{\mathrm{D}} \la 1165$
K. This limit may serve as a new definition for CNM gas; however, the
numerical value is not very well defined from the fit.  It is also not a
sharp limit because the onset of the $f_c(T_{\mathrm{D}})$-correction is
only gradual at $T_{\mathrm{D}} \la 1165$ K. The border between CNM
and LNM was defined previously by \citet[][Fig. 7]{Kalberla2018} as the
Doppler temperature where the frequency distribution for CNM and LNM
Gaussians is equal.  This new limit of 1165 K would be a definition of
the highest temperature where a transition from \hi\ to \h2 is
observable. 

 An excess of $E(B-V) / N_{\mathrm {H}}$ may be affected by
  saturation of the \hi\ emission due to self-absorption or optical
  depth effects. \citet{Fukui2015} investigated these effects and, 
  from the analysis of {\it Planck}/IRAS data toward high galactic
  latitudes, derived 2--2.5 times higher \hi\ densities than under the optical
  thin assumption. They suggested that optically thick \hi\ gas may
  dominate CO--dark gas in the Milky Way. Contrary to their results, we find that  optical
  depth corrections according to Sect. \ref{Initial} item 6   affect the ratio $f_c(T^{\mathrm{sel}}_{\mathrm {D}})$ on
  average by 5\% for a correction according to \citet{Lee2015} and by
  1\% for the correction proposed by \citet{Nguyen2018}. Both
  corrections are too small to explain systematical changes in the
  gas-to-dust ratio.  Similarly, \citet{Liszt2014b} concludes that
  optical depth corrections are too small to have any significant effect
  on the derived $N_{\mathrm {HI}}/ E(B-V)$ ratio.  More recently
  \citet{Murray2018}, using GALFA-\hi\ data, confirm that excess dust
  emission in the local ISM cannot be dominated by optically thick
  \hi\ in the local ISM. \citet{Tang2016} investigated the physical
  properties of CO--dark molecular gas traced by C$^+$. Their sample of
  36 sources close to the Galactic plane should be most sensitive to
  optical depth effects, but they find that the \hi\ optical depth can
  vary in a wide parameter range without significantly affecting the
  global relations between the CO--dark gas fraction $f^N_{H2} = 2 ~
  N_{\mathrm {H2}} / N_{\mathrm {H}}$, and \hi\ excitation
  temperature. \citet{Tang2016} conclude that the molecular gas must be
  the dominant component regardless of individual excitation
  temperatures, optical depth, and the lack of CO emission. Deriving
  excitation temperatures $T_{\mathrm{ex}}$ for an optical depth of 1,
  these authors find a relation $f^N_{H2} = -2.1~ 10^{-3} ~
  T_{\mathrm{ex}} + 1$. This trend, in agreement with results from
  \citet{Rachford2009}, implies $ f_c \propto T^{-1}_{\mathrm{ex}}$,
  broadly consistent with Eq. \ref{eq:f_c}.
  
  Optical depth corrections derived by different groups
  (\citet{Lee2015}, \citet{Murray2018}, and \citet{Nguyen2018}) are
  rather uncertain; we show, that a single parameter dependence of
  $f_c(T_{\mathrm{D}})$-correction according to Eq. \ref{eq:fCsel} is
  sufficient to minimize systematic fluctuations in the $
  E(B-V)/N_{\mathrm {HI}}$ ratio over the full high Galactic latitude
  sky.  Even toward the CO--bright highest HI column density
  star-forming regions, a radiation transfer calculation for \hi\ is
  feasible, which \citet{Li2003} demonstrate in great detail. Our
  analyses are toward the high Galactic latitude sky, away from
  CO--bright or even star-forming regions. Thus, opacity effects are
  avoided by selecting CO--dark regions of the sky. Section
  \ref{Filaments} comprises compelling evidence that toward high
  Galactic latitudes the typical filamentary CNM structures show up with
  low Doppler temperatures $T_{\mathrm{D}} \la 220$ K and FWHM of
  $\delta v_{\mathrm {LSR}} \la 3 $ \kms\ and negligible optical depth
  effects.


\begin{figure}[tbp] 
    \centering
    \includegraphics[width=9cm]{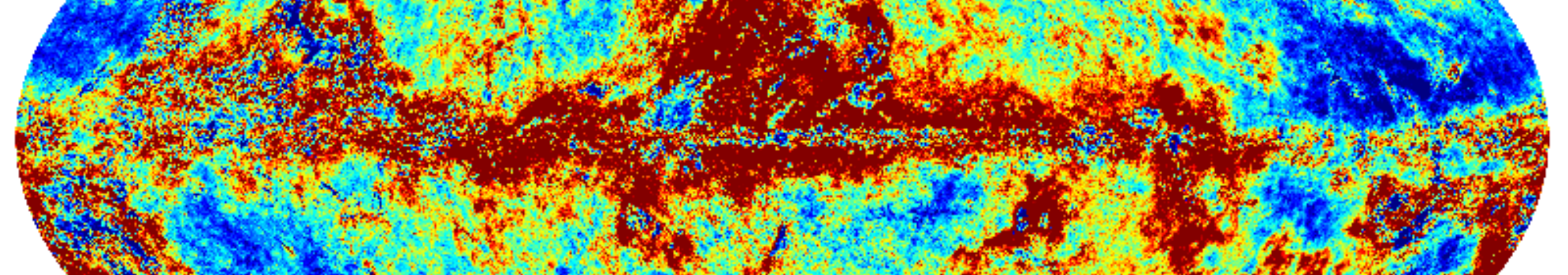}
    \includegraphics[width=9cm]{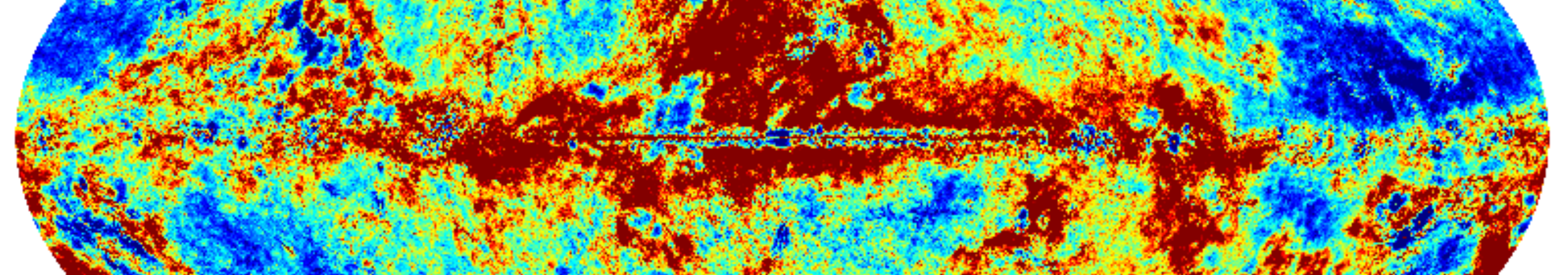}
    \includegraphics[width=9cm]{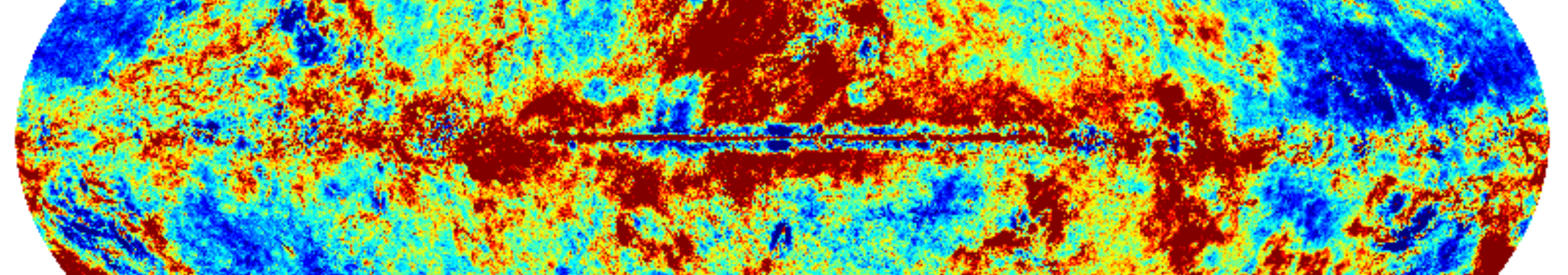}
    \includegraphics[width=9cm]{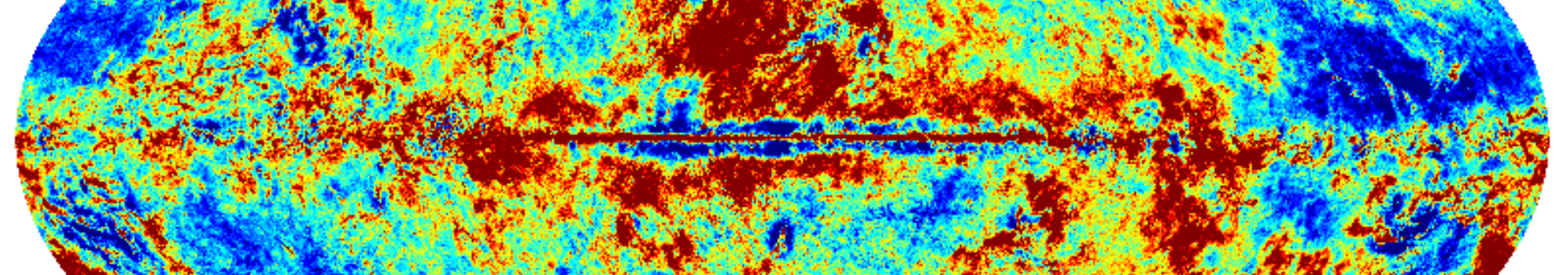}
    \includegraphics[width=9cm]{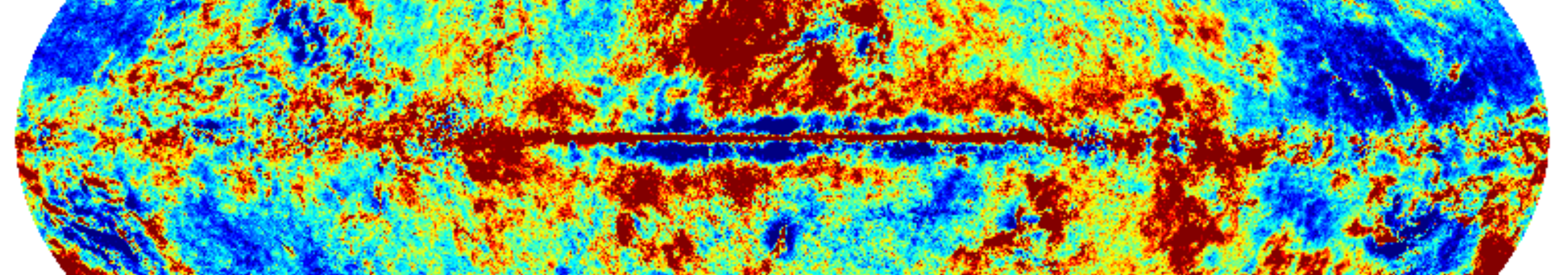}
    \caption{ Galactic plane excerpts for $E(B-V)/N_{\mathrm {Htot}}$
      with a correction for $X_{\mathrm {CO}} = 4.0 \times 10^{20}$
      cm$^{-2}$ (K \kms)$^{-1}$ as in Fig. \ref{Fig_Plot_EB_NH} (bottom
      left), but with different smoothing kernels $S$: (from top to bottom)
      $S$ with FWHM of $ 0\deg, 1\deg, 2\deg, 3\deg, $ and $4\deg$. }
   \label{Fig_Plot_EB_NH_smo}
\end{figure}

\subsection{Spatial distribution of $E(B-V)/N_{\mathrm {H}}$ }
\label{Fits}

We use the $f_c(T_{\mathrm {D}})$ relation according to Eq. \ref{eq:f_c}
to derive an all-sky relationship between gas and interstellar reddening.  We
calculate for each position the ratios $E(B-V)/N_{\mathrm {HI}}$ and
$E(B-V)/N_{\mathrm {H}}$ to generate maps of the spatial distribution of
these ratios. For $E(B-V)/N_{\mathrm {H}}$ we determine an average of
$(1.0 \pm 0.2) ~ 10^{-22}$ cm$^{2}$~mag from a fit of the
$E(B-V)/N_{\mathrm {H}}$ distribution displayed in black in the bottom panel
of Fig. \ref{Fig_EB_NH}.

Figure \ref{Fig_Plot_EB_NH}, left, shows the derived maps.  While in the
top panel only $N_{\mathrm {HI}}$ is accounted for, in the middle panel
the hydrogen column density $ N_{\mathrm {H}}$ is shown according to the
correction Eq. \ref{eq:f_c}. We scale the color-coding to display a
2$\sigma$ range around the average $E(B-V)/N_{\mathrm {H}}$.  The bottom
panel serves as reference; we use the canonical $X_{\rm CO}$ factor to
determine the total column density $N_{\mathrm {Htot}}$ using $^{12}$CO
as tracer of of the CO--bright $N_{\mathrm {H2}}$.  We use CO data from
\citet{Dame2001}\footnote[4]{\url{https://lambda.gsfc.nasa.gov/data/foregrounds/dame_CO/lambda_wco_dht2001.fits}}
to calculate this part of the \h2 distribution but a straightforward
subtraction of the CO--bright $N_{\mathrm {H2}}$ leads to unsatisfactory
results. First we notice that the CO--bright \h2 cannot be determined by
simply using a constant $X_{\mathrm {CO}}$ factor. The second problem is
that a straightforward subtraction of a CO--bright \h2 component leads
to an obvious spatial mismatch between observed and modeled CO--bright
\h2 distribution. Figure \ref{Fig_Plot_EB_NH_smo} can help demonstrate
the problems. From top to bottom we show attempts to model the
CO--bright \h2 distribution with various smoothing kernels.  We aim to
derive the properties of the diffuse CO--dark \h2 but a best possible
fit of the CO--bright \h2 is beyond the scope of this publication.
  
For a reasonable solution we find that the CO data from \citet{Dame2001}
need to be smoothed to a resolution of about 2\deg\
(Fig. \ref{Fig_Plot_EB_NH_smo}, middle). The original survey data were
constructed from several different CO surveys with a grid spacing of
0\fdg125. Surveys with half beamwidth spacing were smoothed with a
Gaussian with FWHM of 0\fdg125. For the lambda data product the original
\citet{Dame2001} data with a resolution ranging from 0\fdg125 to 0\fdg5 were
interpolated to a HEALPix grid with nside = 512, appropriate for a
comparison with $E(B-V)$ data on the same grid. The implication from the
noisy performance of the unsmoothed CO data
(Fig. \ref{Fig_Plot_EB_NH_smo}, top) is that the spatial distributions
of CO and \h2 in CO--bright regions must be different. A significant
part of the \h2 appears to be distributed around dense molecular gas
cores; we refer to the model proposed by \citet{Wolfire2010} and Fig. 1
of \citet{Seifried2020MNRAS}. We apply a factor $X_{\mathrm {CO}} =
4.0 \times 10^{20}$ cm$^{-2}$ (K \kms)$^{-1}$ to calculate the
CO--bright \h2. This factor is high, but still in the range $1.7 <
X_{\mathrm {CO}} < 4.2 \times 10^{20}$ cm$^{-2}$ (K \kms)$^{-1}$ 
determined by several authors from extinction data \citep[][Table
  1]{Bolatto2013}. The bottom left panel of Fig. \ref{Fig_Plot_EB_NH}
shows that such a determination of the CO-associated \h2 leads to an
improved ratio for $E(B-V)/N_{\mathrm {H_{\mathrm {tot}}}}$, but there
are significant deviations from an average dust-to-gas ratio. It is
obvious that the CO--bright \h2 cannot be derived by using a unique
$X_{\mathrm {CO}}$ factor. We conclude that a determination of the
$E(B-V)/N_{\mathrm {H}}$ ratio is currently safe outside CO--bright
regions, but probably also to the full sky with only a few restrictions.

\begin{figure}[thp] 
    \centering
 \includegraphics[width=9cm]{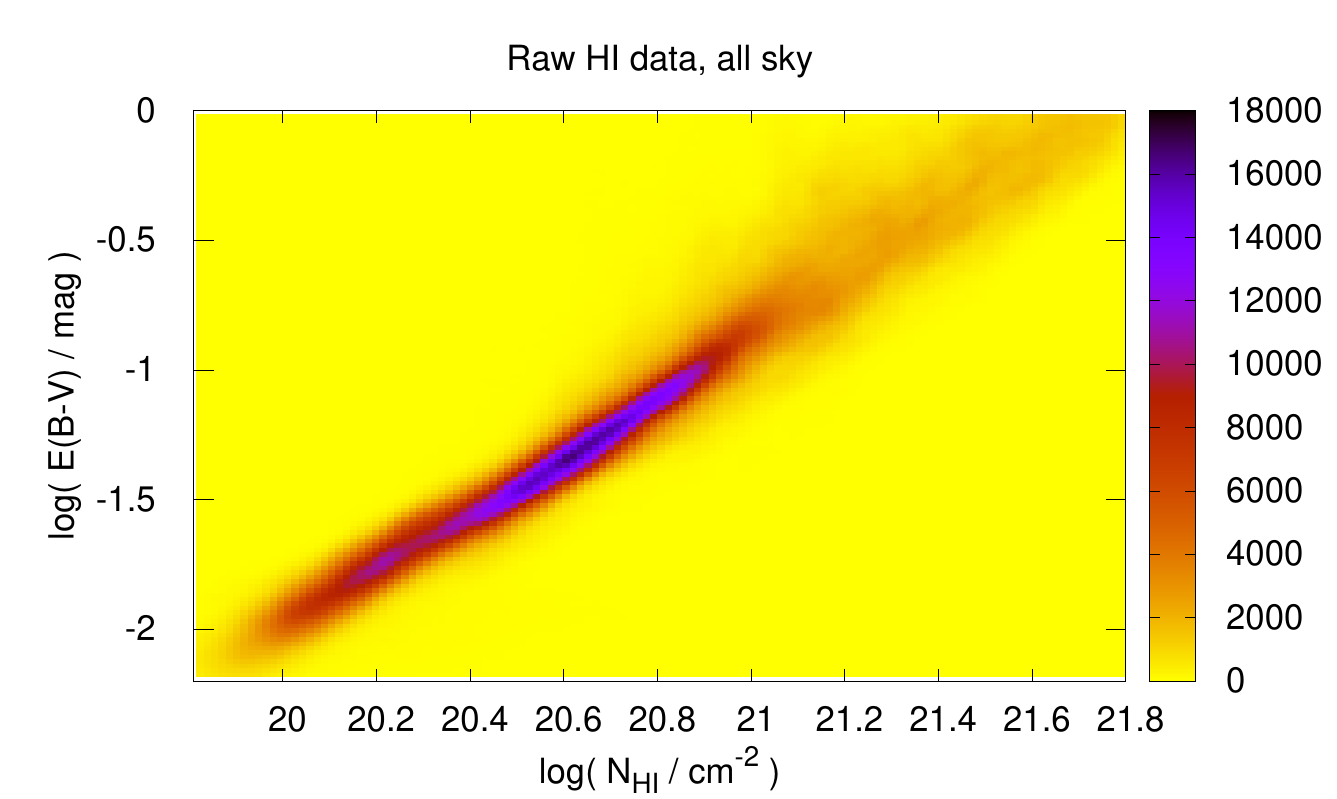}
 \includegraphics[width=9cm]{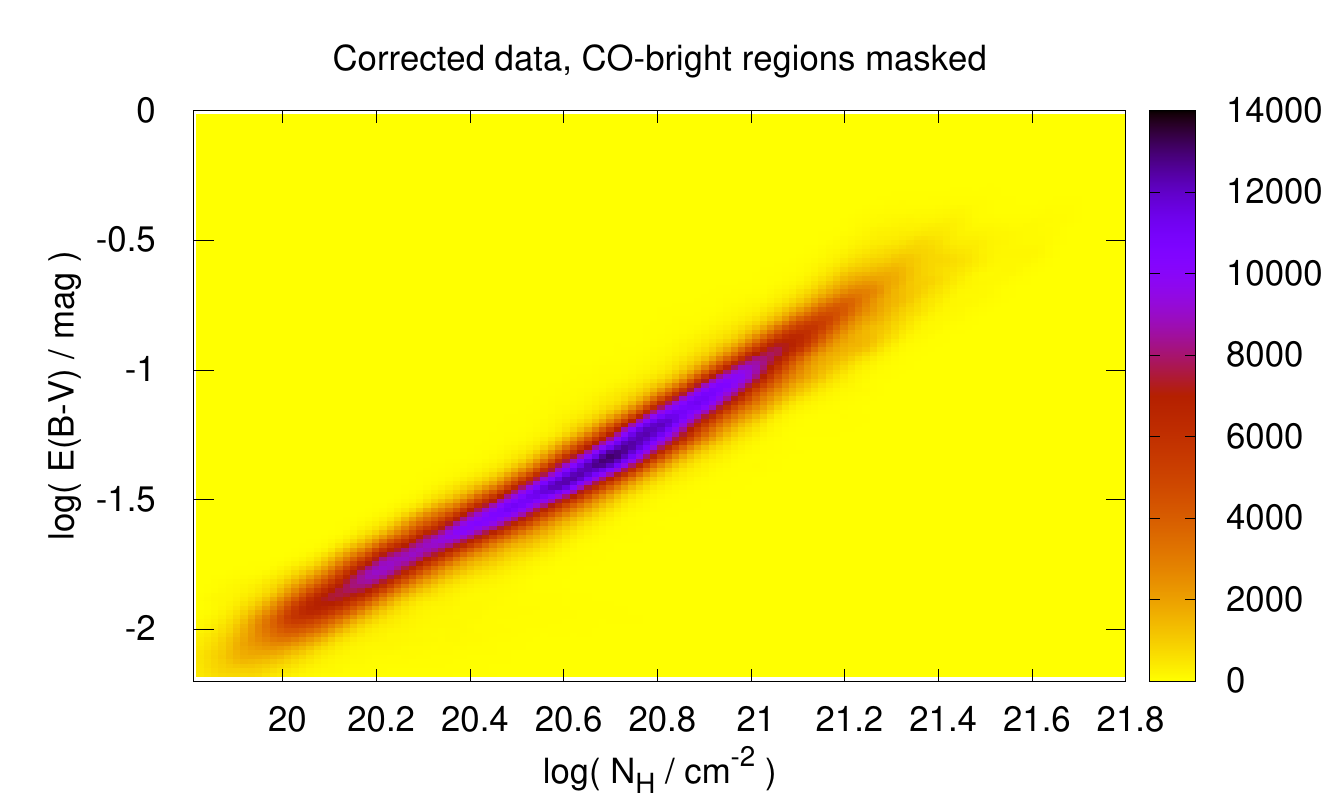}
 \includegraphics[width=9cm]{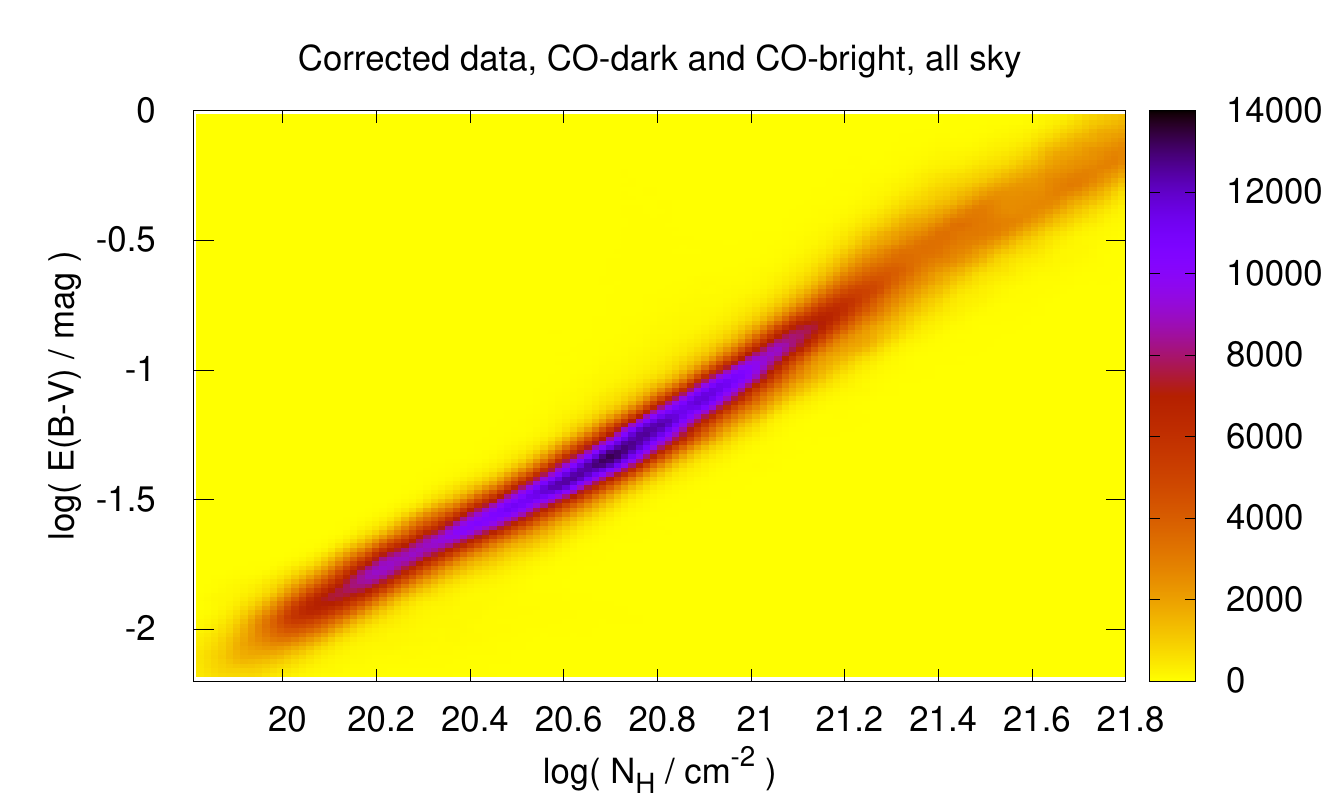}
    \caption{All-sky 2D histograms for the $E(B-V)$ vs. $N_{\mathrm
        {HI}}$ (top) and $N_{\mathrm {H}}$ (bottom) after applying the
      $f_c(T_{\mathrm {D}})$ correction. Middle: Data in CO--bright
      regions were disregarded. The color-coding represents pixel
      counts. }
   \label{Fig_Plot_regres}
\end{figure}

An independent estimate for the FIR opacity caused by the dust is given
by {\it Planck} maps of the optical depth $\tau_{353}$ at 353 GHz
(\citet{Planck2011} and \citet{Planck2016}). We used the optical depth
map from the \citet{Planck2016} data
release\footnote[5]{\url{http://pla.esac.esa.int/pla/aio/product-action?MAP.MAP_ID=COM_CompMap_Dust-GNILC-Model-Opacity_2048_R2.00.fits}}
and calculated the ratios $\tau_{353}/N_{\mathrm {HI}}$,
$\tau_{353}/N_{\mathrm {H}}$, and $\tau_{353}/N_{\mathrm {Htot}}$. The
results are shown on the right  side of
Fig. \ref{Fig_Plot_EB_NH}. We used again a scaling of the color-coding such
that a 2$\sigma$ range around the average $\tau_{353}/N_{\mathrm {H}}$
is displayed. We determine an average of $(7.10 \pm 1.7 ) ~ 10^{-27} $
cm$^{2}$. The maps in the left and right panels of
Fig. \ref{Fig_Plot_EB_NH} should be comparable, but we find some striking
large-scale differences, existing both for raw and corrected \hi\ data,
indicating that there are unaccounted for systematic uncertainties.

\begin{figure*}[tbp] 
    \centering
    \includegraphics[width=9cm]{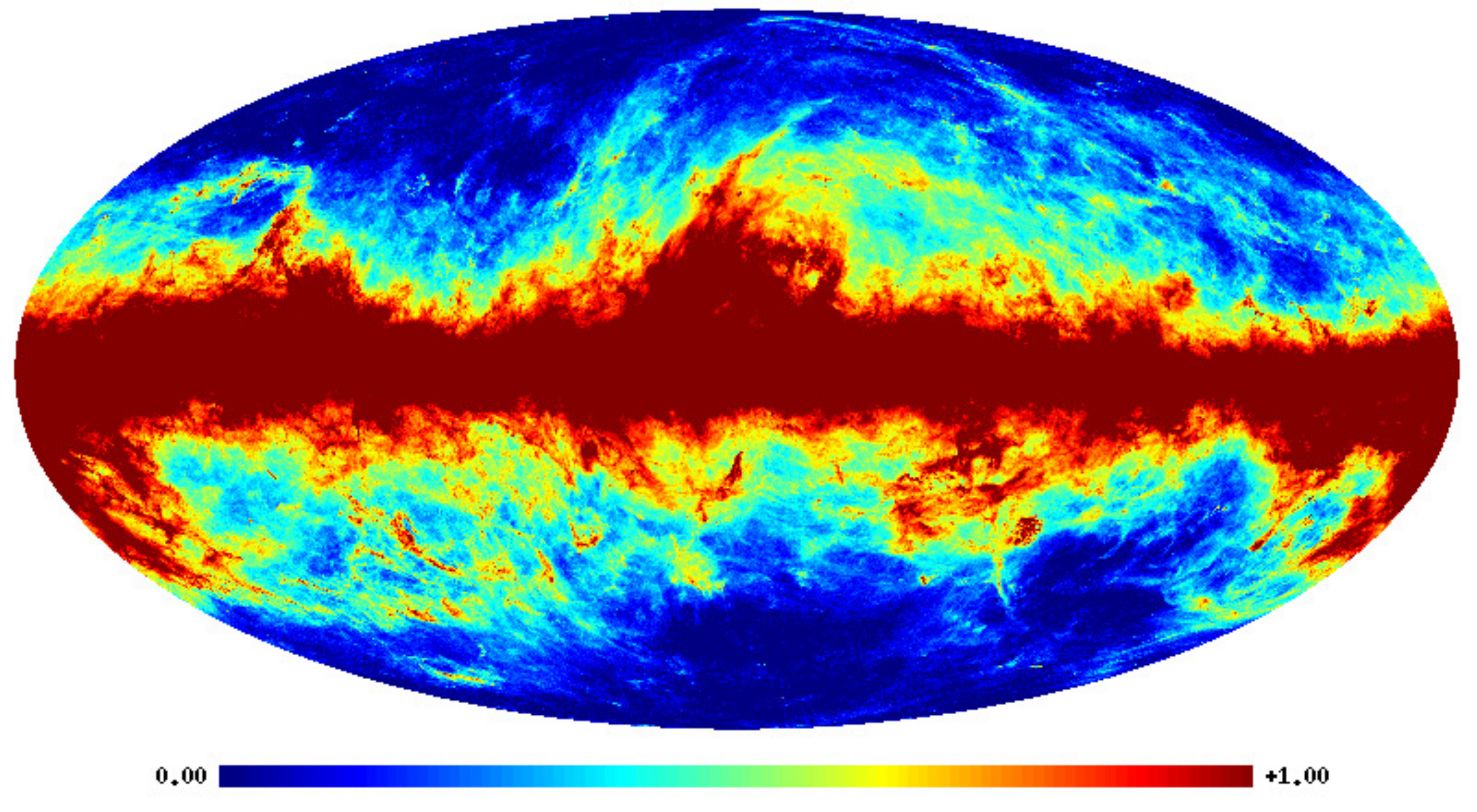}
    \includegraphics[width=9cm]{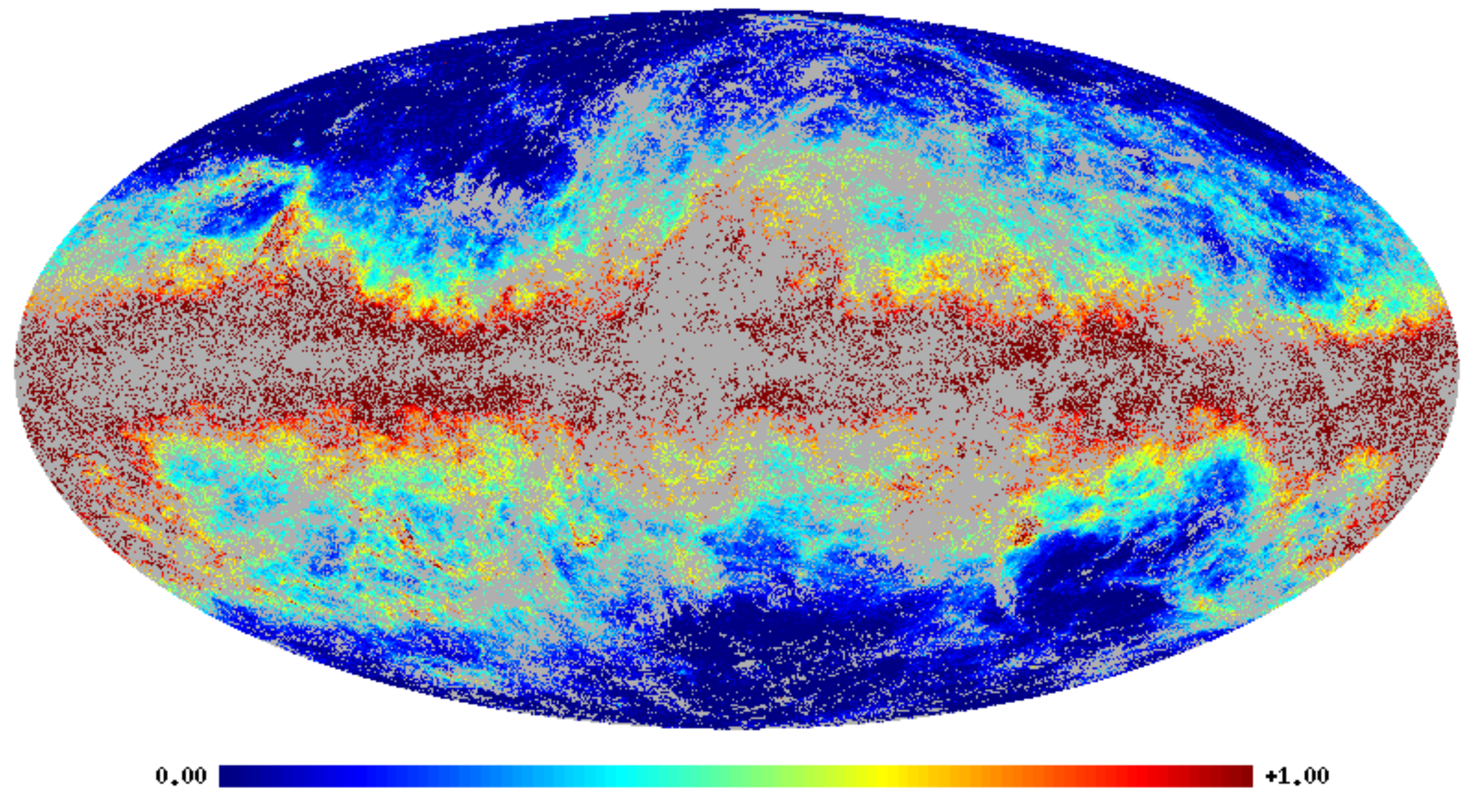}
    \includegraphics[width=9cm]{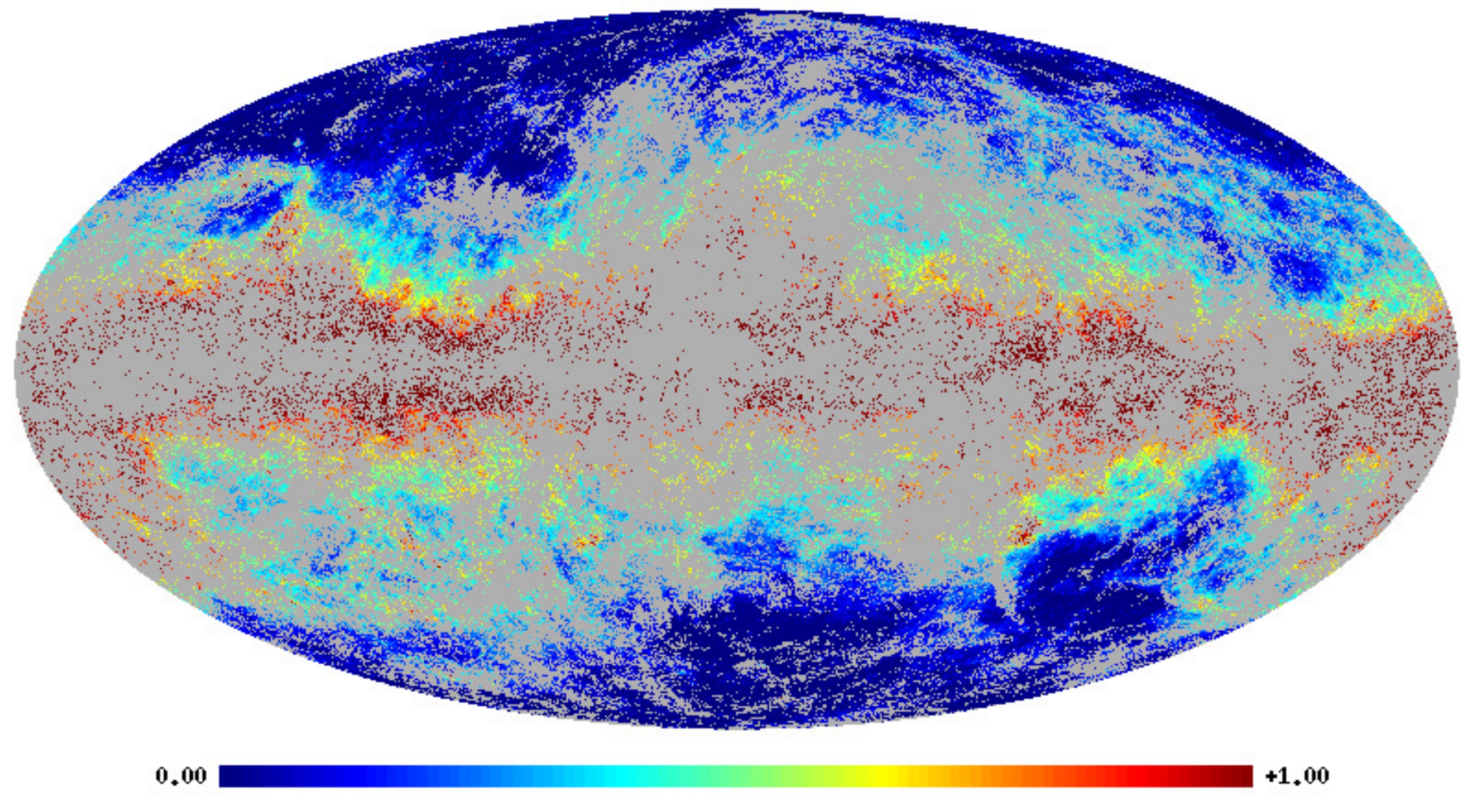}
    \includegraphics[width=9cm]{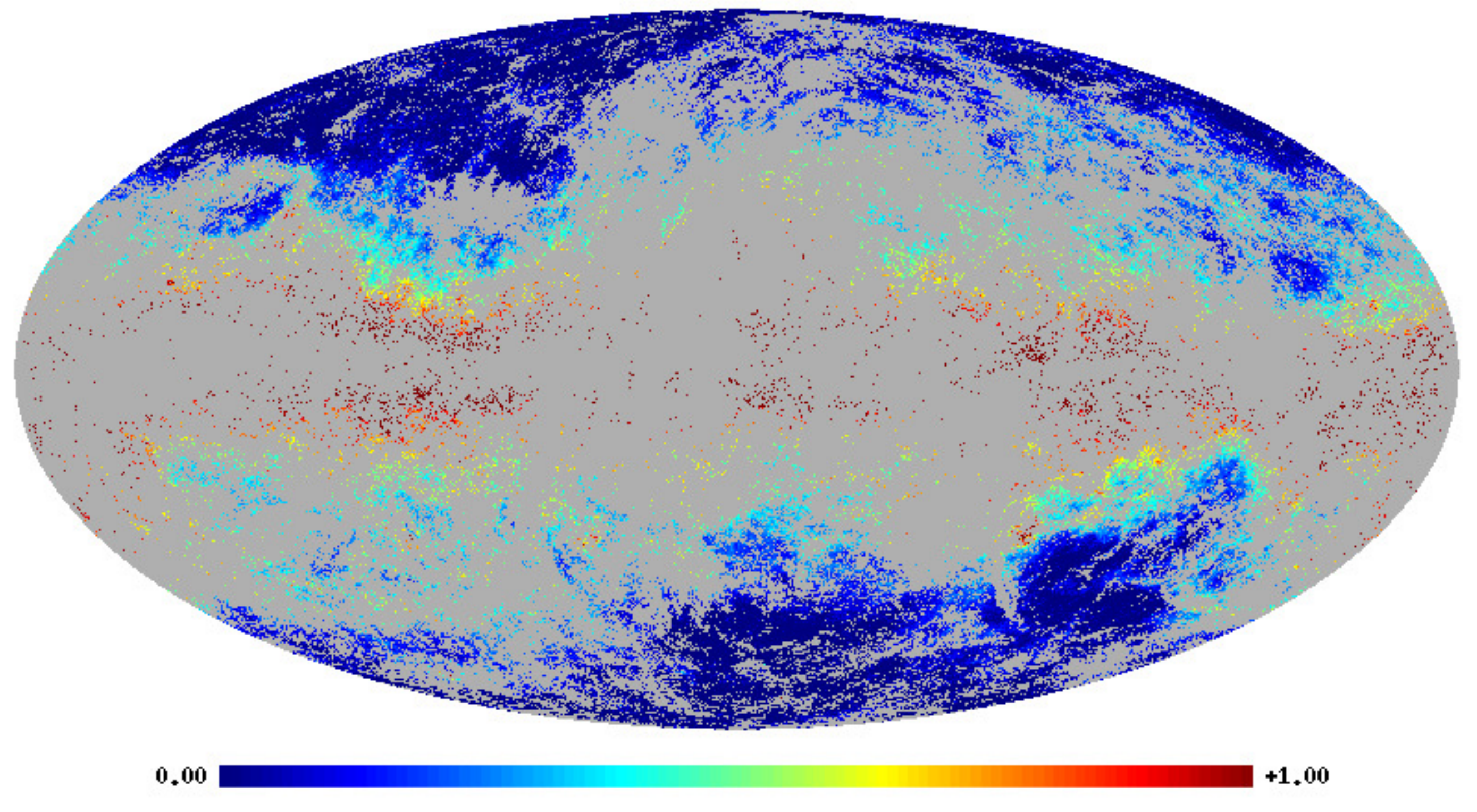}
    \caption{ Far-IR emission at 857 GHz as observed by {\it Planck},
        original data at top left. To indicate upper limits for
        $T_{\mathrm {kin}}$, positions with CNM Doppler temperatures
        below thresholds of 155, 220, and 311 K (top right, bottom left,
        and bottom right) are masked.  The color scale represents
        log($I_{857}$/MJy sr$^{-1}$). } 
   \label{Fig_SkyMap_857}
\end{figure*}

Column densities for individual positions from \hi\ surveys have typical
uncertainties on a 2.5\% level \citep{Winkel2016}, but there are no
additional uncertainties on large scales.  We conclude that at least
some of the systematic deviations from the mean must come from the
extinction data. For the dust-to-gas ratio $E(B-V)/N_{\mathrm {H}}$
using the data from \citet{Schlegel1998} we derive outside CO--bright
regions a relative scatter of 21\%. This compares to a scatter of 24\%
for $\tau_{353}/N_{\mathrm {H}}$. According to \citet{Schlegel1998} some
of the $E(B-V)$ data are known to contain fluctuations of $\pm 15$\%
amplitude that are coherent over scales of $\ 10\deg$. These may be real
variations from gas to dust, or they may trace some unresolved instrumental
or systematical problems such as shortcomings in the temperature
corrections \citep{Schlafly2010}. For a comparison between $E(B-V)$ and
$\tau_{353}$ we refer to \citep[][Sect. 7.3, Fig. 26]{Planck2014}.  At a
resolution of 6\farcm1 and for $N_{\mathrm {HI}} < 2~ 10^{20} $
cm$^{-2}$, the ratio $E(B-V)/\tau_{353}$ is constant within a scatter of
7\%. Smoothing both data sets to a resolution of 30\arcmin\ the ratio
$E(B-V)/\tau_{353}$ shows local variations larger than 30\% all over the
sky. This database is no longer available for download. Using the
optical depth map from \citet{Planck2016},  currently available in the
official distribution (see footnote 5), we obtain an all-sky   rms scatter
of 23\% for the nside = 1024 databases  and 20\% after
30\arcmin\ smoothing.

Systematical large-scale deviations from the average dust-to-gas ratio
cause a significant fraction of the scatter of $R_{\mathrm
  {HI}}(T^{\mathrm{sel}}_{\mathrm D})$ and $R_{\mathrm
  {H}}(T^{\mathrm{sel}}_{\mathrm D})$ in Fig. \ref{Fig_EB_NH} and also
uncertainties in fitting an $f_c(T_{\mathrm{D}})$ correction to the
\hi\ data. It is hard to estimate how far regression parameters in
Eq. \ref{eq:f_c} are affected by such systematical problems. Our
estimate of the CO--bright \h2 distribution is strongly affected by
remaining enhancements in the gas-to-dust ratio near  the Galactic
plane, see Fig. \ref{Fig_Plot_EB_NH} bottom. Here we have an additional
problem with variations caused by expected uncertainties in $X_{\mathrm
  {CO}}$. We decided not to apply a fit to large-scale enhancements in
$E(B-V)/N_{\mathrm {Htot}}$.

\subsection{The $N_{\mathrm {H}}/E(B-V)$
  ratio after $f_c(T_{\mathrm {D}})$ correction }
\label{N_H_EB}

Dependences of the \hi\ determined $N_{\mathrm
  {H}}/E(B-V)$ ratio on various selection criteria have
been discussed in great detail by \citet{Liszt2014a,Liszt2014b}, and by
\citet{Lenz2017} who concluded that it is possible to derive any
$N_{\mathrm {HI}}/E(B-V)$ value between
those of \citet{Bohlin1978} and \citet{Liszt2014b} depending simply on
the range of column densities comprised by the fit. In the following we
want to check whether this situation has improved after application of
the $f_c(T_{\mathrm {D}})$ correction.

We calculate all-sky  ratios corresponding to the panels on the
left  side of Fig. \ref{Fig_Plot_EB_NH}. In the  top panel  of
Fig. \ref{Fig_Plot_regres} we display a 2D histogram of $N_{\mathrm
  {HI}}/E(B-V) $ from HI4PI as observed. In the middle panel we show the
distribution of $N_{\mathrm {H}}/E(B-V) $ after application of
$f_c(T_{\mathrm {D}})$. We mask CO--bright regions, thus this histogram
is valid for all diffuse \h2 regions that are not affected by additional
\h2 that might be associated with CO. In the bottom panel of Fig.
\ref{Fig_Plot_regres} we show a 2D histogram using all-sky \hi\ and
CO--dark as well as estimated CO--bright \h2, as shown in
Fig. \ref{Fig_Plot_EB_NH}, bottom left.

The 2D histograms in logarithmic scale from Fig. \ref{Fig_Plot_regres}
can be directly compared to Fig. 1, bottom right
panel, in  \citet{Lenz2017}. Remarkable is the absence of the increase in scatter reported by
\citet{Lenz2017} above $N_{\rm HI} \geq 4 ~10^{20} \mathrm {cm}^{-2}$ in
our corrected data. The maximum  number of components is around
$N_{\rm HI} \simeq 5 \times 10^{20}\,{\rm cm^{-2}}$, and here we see a
slight bending in the gas-to-dust ratio. Even above this value the
linear correlation remains very well defined.

Using all-sky data without any constraints we obtain $N_{\mathrm {H}}/
E(B-V) \sim 5.1 ~ 10^{21} \mathrm {cm}^{-2} \mathrm
{mag}^{-1} $.  Excluding latitudes $|b| \le 8\deg$ we get
$N_{\mathrm {H}}/ E(B-V) \sim 6.7 ~
10^{21} \mathrm {cm}^{-2} \mathrm {mag}^{-1} $.  Both results are
affected by systematic errors, but bracket the previous determination
$N_{\mathrm {H}}/ E(B-V) \sim 5.8 ~
10^{21} \mathrm {cm}^{-2} \mathrm {mag}^{-1} $ by
\citet{Bohlin1978}. They are also consistent with the more recent
determination $N_{\mathrm {H}}/E(B-V) =
(6.45 \pm 0.06) ~ 10^{21} \mathrm {cm}^{-2} \mathrm {mag}^{-1} $ by
\citet{Zhu2017} who considered X-ray observations of a large sample of
Galactic sightlines.

\section{The nature of \hi\ filaments}
\label{Filaments}  

Data presented by \hi\ observers led recently to a picture of the
neutral ISM with the CNM distributed into cold, small-scale anisotropic
structures, preferentially aligned along the magnetic field and
associated with dust (e.g., \citet{Heiles2005}, \citet{Clark2014},
\citet{Kalberla2016}, \citet{Clark2019}, and \citet{Kalberla2020}).
Supporting evidence for low temperatures at the position of filaments
was recently reported from ${\ion{Na}{i}}$ absorption measurements of
50,985 quasar spectra by \citet{Peek2019}.  The interpretation that
filamentary structures are cold is  questioned
frequently, however. Small-scale structure in \hi\ channel maps is often assumed
to originate from velocity caustics, caused by the turbulent velocity
field, rather than from real density structures
(e.g., \citet{Lazarian2000}, \citet{Lazarian2018}, and \citet{Yuen2019}).
In this context we need to discuss the structure of \hi\ in filaments,
in particular the distribution of temperatures and velocities and their
relation to FIR emission.

\subsection{$T_{\mathrm {D}}$ along the bones of \hi\ filaments }
\label{bones}

We use Doppler temperature thresholds to demonstrate their response to
FIR emission observed with {\it Planck} at 857
GHz\footnote[6]{\url{https://pla.esac.esa.int/pla-sl/data-action?MAP.MAP_OID=14628}}. The
median Doppler temperature of the CNM at high Galactic latitudes is 220
K (\citet{Clark2014}, \citet{Kalberla2016}, and \citet{Kalberla2019}). We
use this value to mask the observed 857 GHz emission at each position
where an \hi\ Gaussian component with a Doppler temperature below this
temperature threshold is found. The masking is repeated by changing the
temperature threshold by a factor of $\sqrt{2}$, hence we use upper
$T_{\mathrm {D}}$ limits of 155, 220, and 311 K. Results from this
masking are shown in Fig. \ref{Fig_SkyMap_857}. For a threshold of 155 K
only prominent filaments are masked; a threshold of 220 K affects most
of the filaments, and with an upper limit of 311 K just a few weak and
diffuse filaments survive the masking. Thus, cold filamentary CNM
structures at high Galactic latitudes mark enhanced FIR emission at 857
GHz; lower $T_{\mathrm {D}}$ values are observed at the position of the
most pronounced filaments. Filamentary structures get more diffuse at
higher $T_{\mathrm {D}}$ values, implying a 3D structure with lowest
temperatures at the bones (or centers) of the filaments within a warmer
and   more diffuse environment. These results  probably imply that most
of the filamentary \hi\ structures are caused by fibers
\citep{Clark2014} rather than sheets as advocated previously by
\citep{Heiles2005C}.

\begin{figure}[thp] 
    \centering
 \includegraphics[width=9cm]{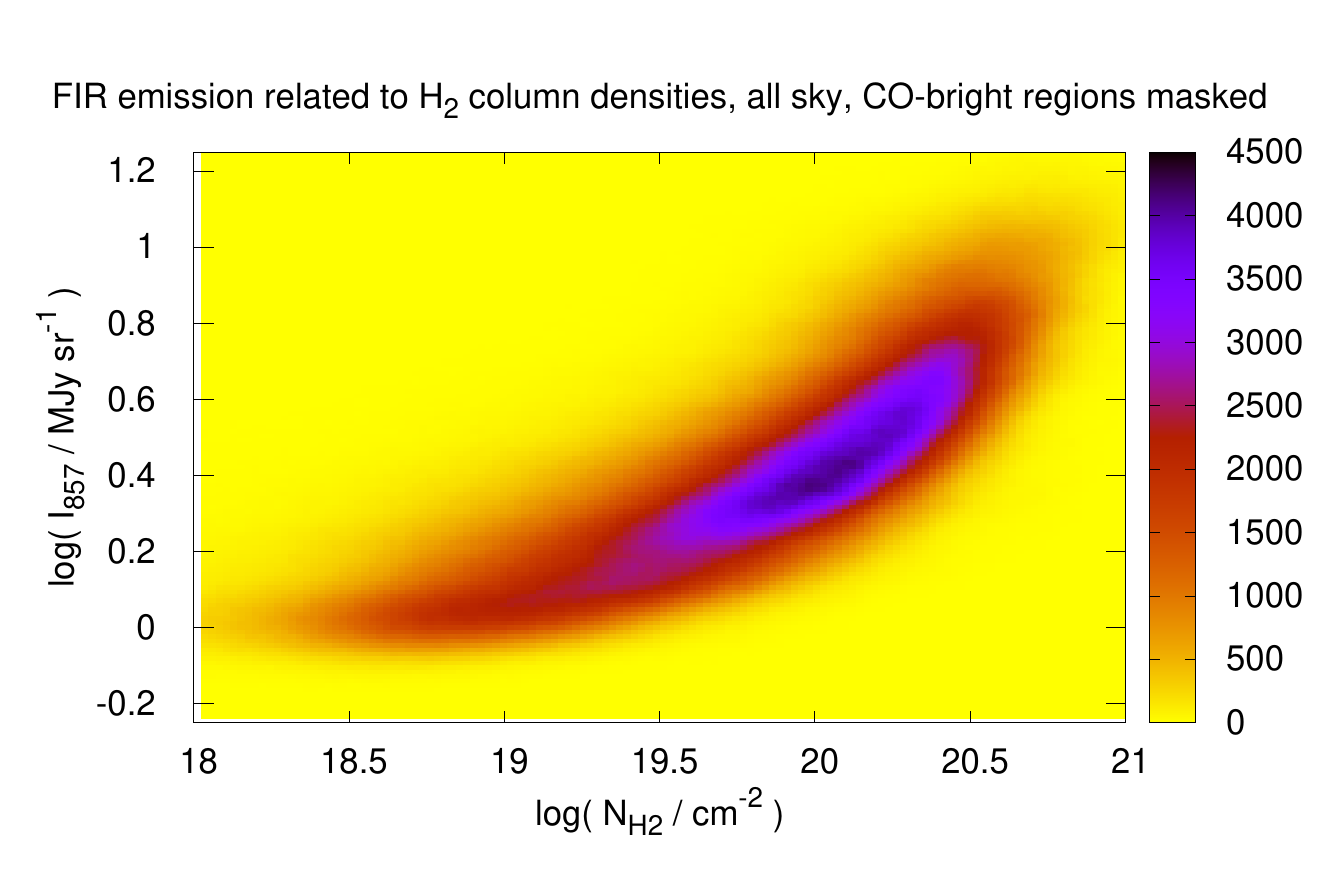}
 \caption{ Histogram of the 2D density distribution of FIR emission
     intensities at 857 GHz and \h2 column densities for $T_{\mathrm{D}}
     < 1165$ K outside CO--bright regions, covering 65\% of the sky. The
     color-coding represents pixel counts. } 
   \label{Fig_Histo_filaments}
\end{figure}

\subsection{Far-IR emission from cold \hi\ filaments }
\label{FIR}

The masking shown in Fig. \ref{Fig_SkyMap_857} can be repeated for dust
emission observed with {\it Planck} at other frequencies and
demonstrates unambiguously that CNM and FIR filaments are associated
with each other.  We interpret this correlation as an indication for the
presence of diffuse CO--dark \h2. Figure \ref{Fig_Histo_filaments}
displays a 2D histogram derived for the distribution of CO--dark \h2
according to Eq. \ref{eq:f_c} and FIR emission in filaments at 857
GHz. There is a clear trend of FIR intensities $I_{857}$ increasing
progressively with increasing \h2 column densities. It was noted
previously by \citet{Clark2019} and \citet{Kalberla2020} that the
$I_{857}/N_{HI}$ ratio increases significantly with the intensity of
\hi\ small-scale structures. We find that the \h2 is correlated with the
CNM (Eq. \ref{eq:f_c}) and at the same time with dust
(Figs. \ref{Fig_SkyMap_857} and \ref{Fig_Histo_filaments}). This way
cold \hi\ is linked to dust filaments, but only for $T_{\mathrm {D}} <
1165$ K, in the presence of diffuse \h2. Filamentary structures associated
with warmer \hi\ are not observed (see Fig 13 in
\citealt{Kalberla2016}).

\begin{figure}[thp] 
    \centering
 \includegraphics[width=9cm]{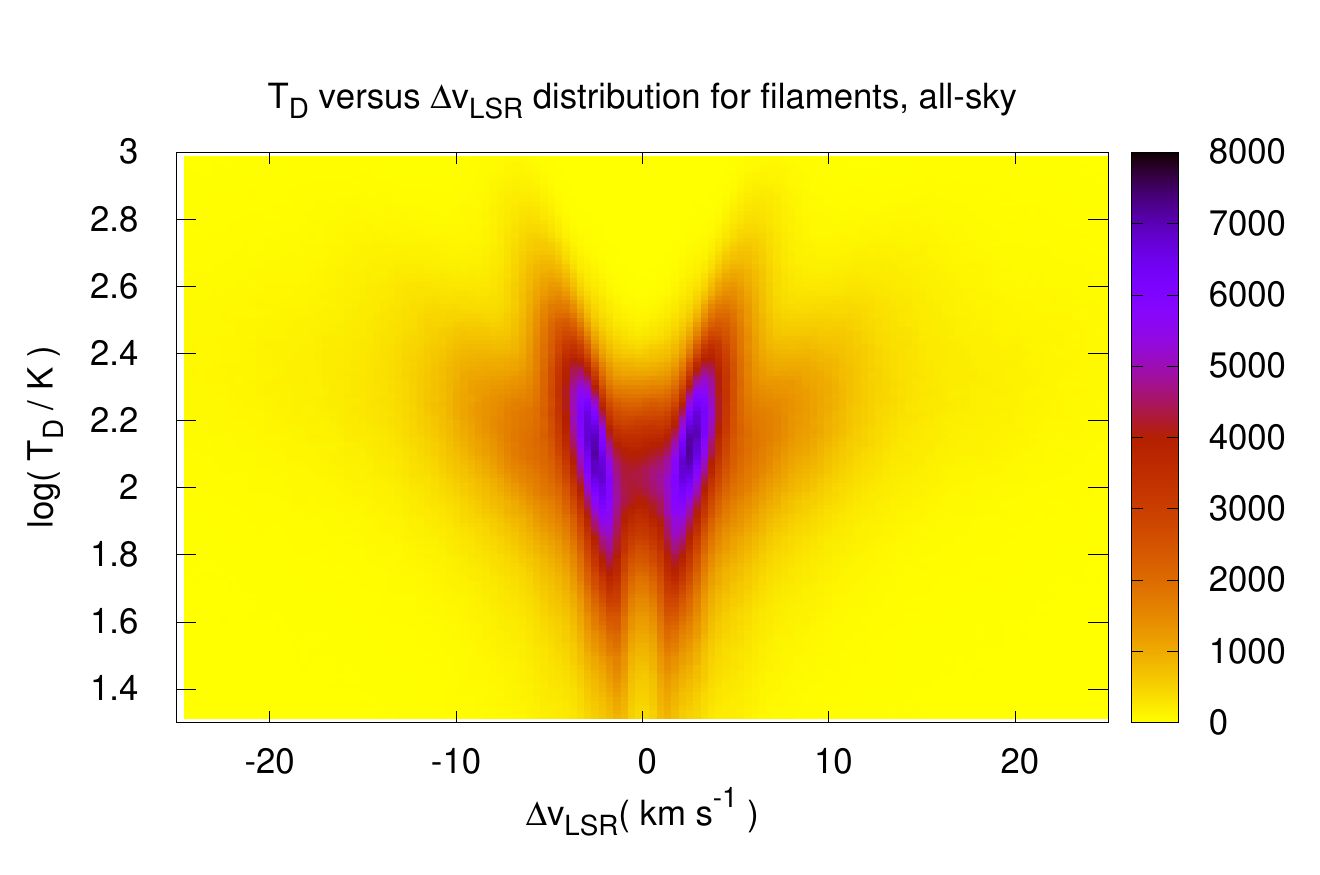}
 \caption{ Histogram of the 2D density distribution of the internal
     velocity structure perpendicular to the bones of
     \hi\ filaments. $\Delta v_{\mathrm {LSR}}$ is defined in the
     velocity domain as the deviation of the nearest Gaussian component
     relative to the component with the lowest $T_{\mathrm {D}}$. }  
   \label{Fig_Histo_vel}
\end{figure}

\subsection{Internal velocity structure of \hi\ filaments }
\label{vel}

The majority of the structures visible in Fig. \ref{Fig_SkyMap_857} are
located in the plane of the sky, otherwise we would not be able to
recognize their filamentary structure. \citet{Clark2018} and
\citet{Clark2019b} derived 3D Stokes parameter maps to constrain
the coherence of these 3D structures and the orientation of the filaments
and their relation to the interstellar magnetic field.  Figure
\ref{Fig_SkyMap_857} suggests that dust, giving rise to FIR emission,
and \h2 as the coldest constituents are located in the centers (the
bones) of the filaments. The warmer \hi\ appears to be distributed
around the bones.

To characterize this situation we use the simplified model of
\hi\ distributed as a tube, encapsulating dust and \h2. In the case of
caustics (\citet{Lazarian2000}, \citet{Lazarian2018}, and
\citet{Yuen2019}) velocities along the line of sight are constant.  If however
the filaments are  density structures that are not exactly in
dynamical equilibrium, we may have the chance to observe two boundary
layers around the central bone containing \h2 and dust. We check our
Gaussian data by searching at each position for two narrow Gaussian
components that might be related to each other. First we determine the
velocity $v_0$ for the component with the lowest Doppler temperature
$T_{\mathrm {D}}$. Next we search at the same position for the Gaussian
component with the lowest velocity deviation $ |v_1 - v_0 | $ that
satisfies the condition $T_{\mathrm {D}} < 1165$ K. $\Delta v_{\mathrm
  {LSR}} = v_1 - v_0$ then defines  the velocity difference between the two
boundary layers or skins along the line of sight.

For 54\% of all positions we obtain two closely related narrow Gaussian
components. Our results are displayed in Fig. \ref{Fig_Histo_vel}. The
2D histogram shows a highly symmetric butterfly diagram with well-defined velocity differences indicating pairs of closely related CNM
components along the line of sight. For $T_{\mathrm {D}} > 85 $ K we fit
$ | \Delta v_{\mathrm {LSR}} | = -6.1 + 3.8 \times \log (T_{\mathrm
  {D}}) $. At a characteristic Doppler temperature $T_{\mathrm {D}} \sim
220$ K this implies a component separation $ \Delta v_{\mathrm {LSR}}
\sim 2.8 $ \kms. In comparison to $ \delta v_{\mathrm {LSR}} = 3.17 $
\kms, the velocity width of the Gaussian with $T_{\mathrm {D}} = 220$ K,
this velocity shift is sufficient to exclude errors from line
blending. Also for other values $T_{\mathrm {D}}\ga 85 $ K line blending
is unimportant. The distribution of the $ \Delta v_{\mathrm {LSR}} $
values on the sky is mostly random without preference for positive or
negative $ \Delta v_{\mathrm {LSR}} $ values.

Interpreting $ \Delta v_{\mathrm {LSR}} $ as being caused by turbulent motions,
we can deconvolve for the line broadening to estimate the excitation
temperature as $ T_{\mathrm{ex}} = 21.86 \times (\delta v^2_{\mathrm
  {LSR}} - \Delta v^2_{\mathrm {LSR}} ) $. For a median Doppler
temperature $T_{\mathrm {D}} \sim 220 $ K we obtain the characteristic
excitation  temperature $ T_{\mathrm{ex}} \sim 48.5$ K. Similar for
$T_{\mathrm {D}} \sim 155 $ K filaments (Fig. \ref{Fig_SkyMap_857}, top
right) we get $ T_{\mathrm{ex}} \sim 47$ K. These estimates are in
excellent agreement with the median $ T_{\mathrm{ex}} \sim 50$ K
derived by \citet{Heiles2005}.  We conclude that the \hi\ filaments are
cold and at least some of this CNM is encapsulating the \h2, consistent
with the finding in Sect. \ref{bones} that filaments are 3D structures
with the lowest temperatures interior at the bones. 

\section{Discussion}
\label{Discussion}

\begin{figure*}[tbp] 
    \centering
    \includegraphics[width=9cm]{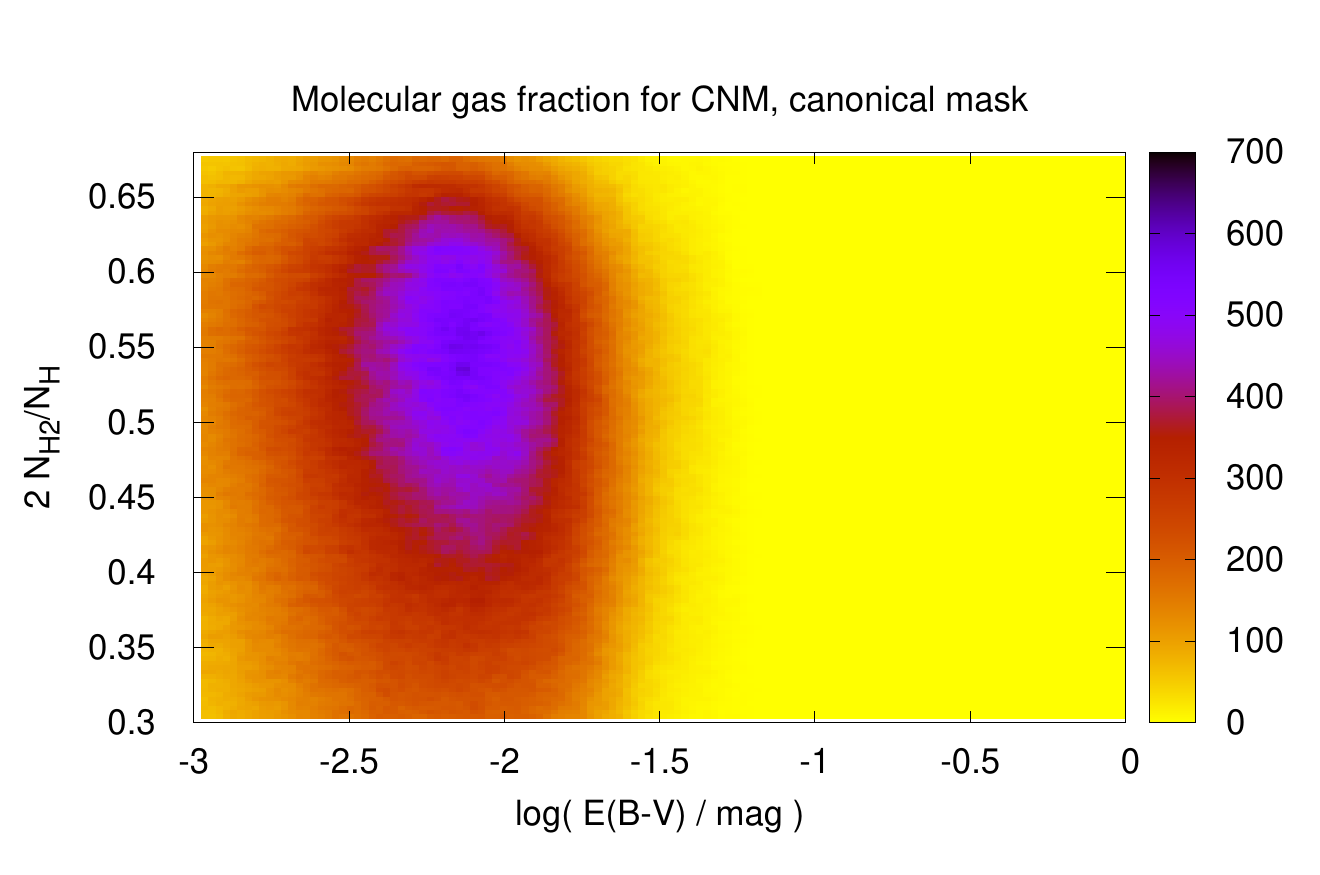}
    \includegraphics[width=9cm]{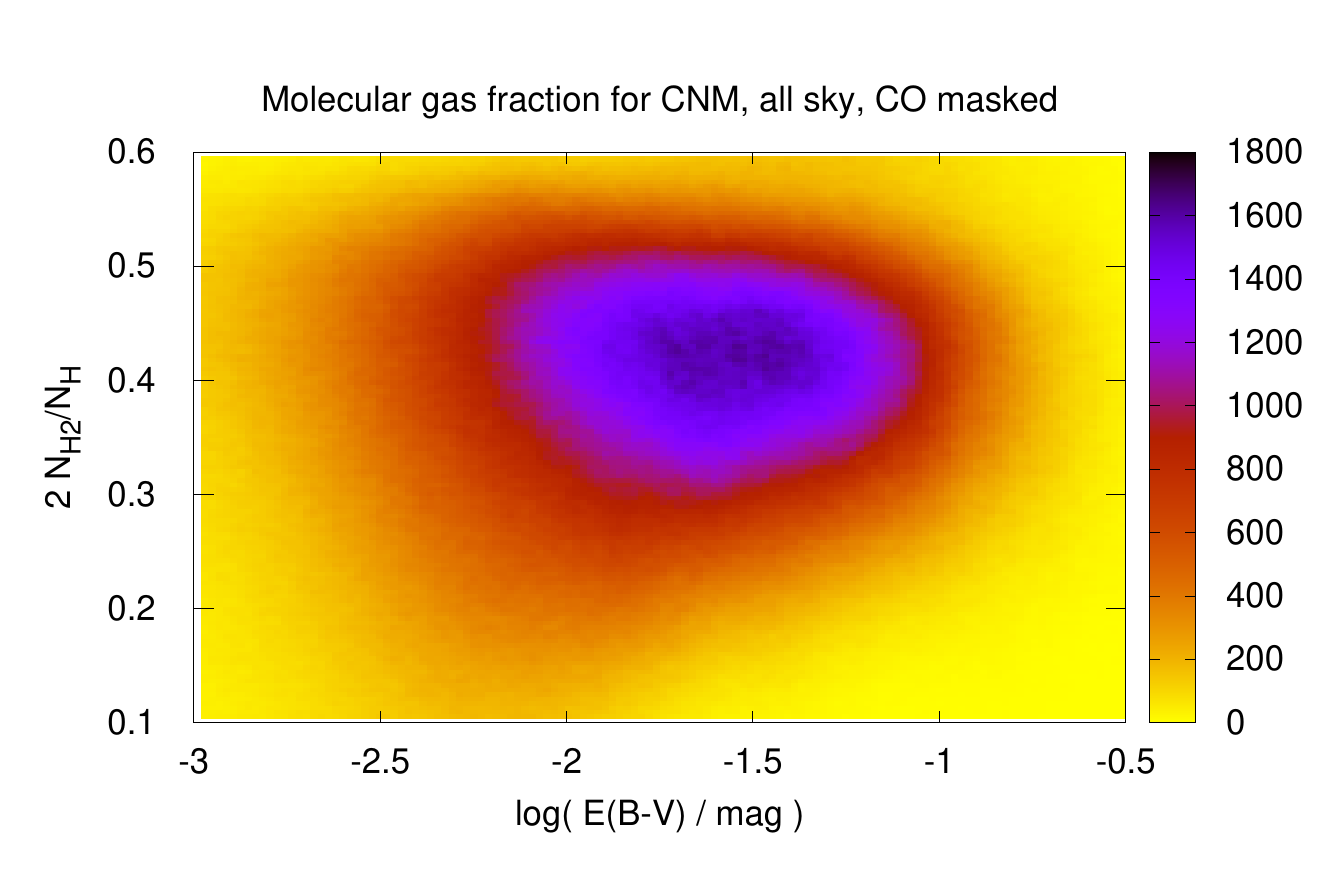}
    \includegraphics[width=9cm]{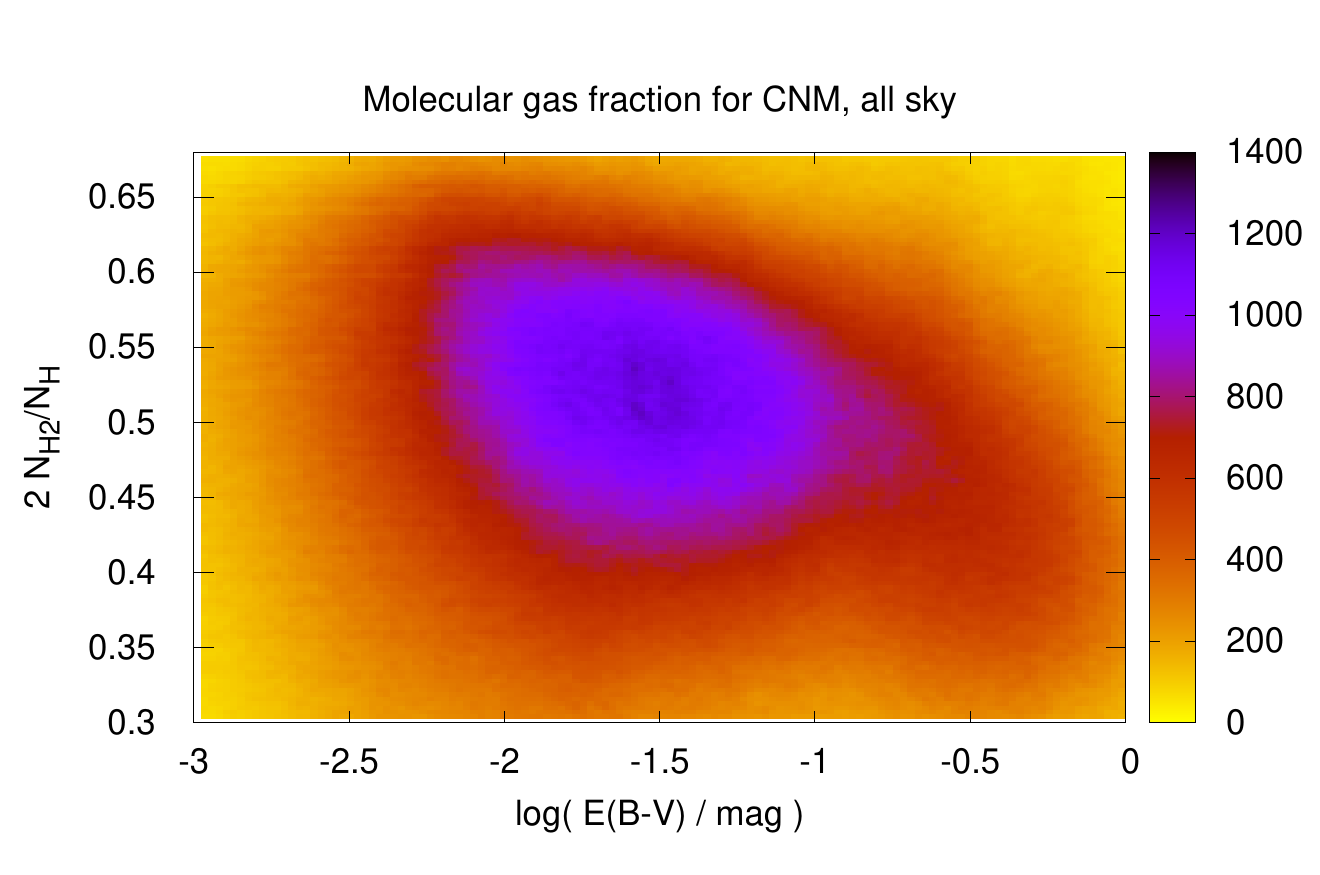}
    \includegraphics[width=9cm]{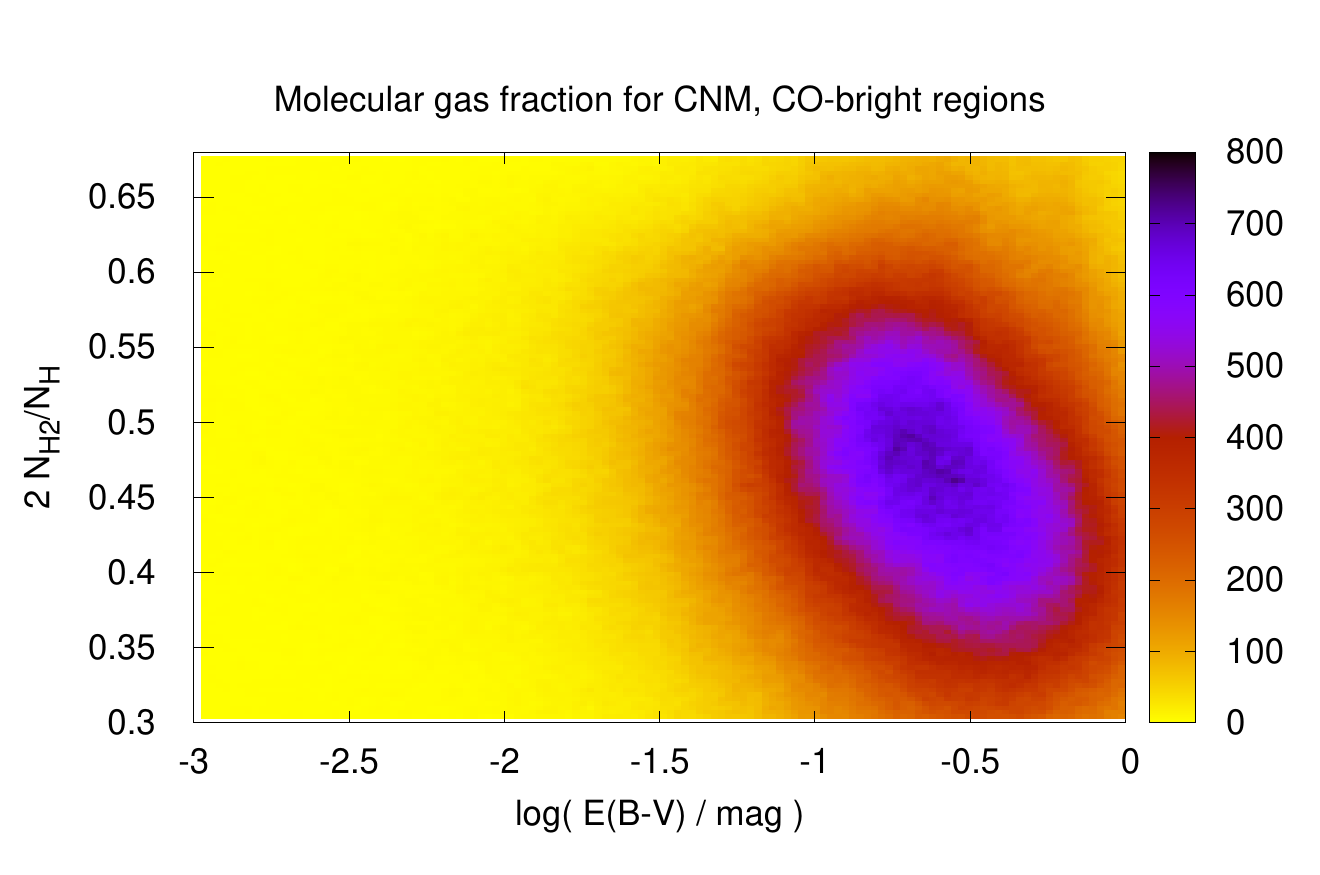}
    \caption{2D histograms of the CNM molecular gas fractions
      $f^N_{\mathrm {H2}} = 2~N_{\mathrm {H2}} / N_{\mathrm {H}} = 1 -
      f^{-1}_c $ depending on $E(B-V)$. Selection criteria are for
        top left: $N_{\mathrm {HI}} \le 4 ~ 10^{20}$ cm$^{-2}$ and
        $E(B-V) \le 0.08$ mag, top right: all-sky but CO regions masked,
        bottom left: all-sky, bottom right: CO--bright regions only. The
        color coding represents pixel counts.} 
   \label{Fig_Plot_2D_hist}
\end{figure*}


In Sect. \ref{Initial} we apply some commonly adapted assumptions to
determine the initial estimates to solve Eq. \ref{eq:EB}.  After
quantifying $f_c(T_{\mathrm {D}})$ iteratively it became apparent, that
these standard assumptions can be released without losing the statistical
significance of the results. More important is our finding that after the 
application of a $f_c(T_{\mathrm {D}})$ correction according to
Eq. \ref{eq:f_c} the commonly applied restriction to high latitudes or
to regions with $E(B-V)/N_{\mathrm {HI}}$ and $N_{\mathrm {HI}} \la 4 ~
10^{20}$ cm$^{-2}$ and $E(B-V) \la 0.08$ mag can be released.  The
$f_c(T_{\mathrm {D}})$ correction is also applicable toward regions with
significant amounts of \h2. However, we need to exclude those regions
with bright CO emission from the fitting.  Inferring the molecular gas
contribution from \hi\ in regions with CO emission does not lead to
a reliable presentation of the data without considering additional \h2
associated with CO in these regions (Fig. \ref{Fig_mask} bottom). As a
result the $f_c(T_{\mathrm {D}})$ correction is applicable to at least
74\% of the sky, quite an improvement to the validity limitations of
the previous \hi-based determinations of the dust-to-gas relation
\citep{Lenz2017}. We find that only 8.8\% of the sky are unaffected by
the $f_c(T_{\mathrm {D}})$ correction. We also tried  to determine the
\h2 distribution within CO--bright regions. We apply a factor
$X_{\mathrm {CO}} = 4.0 \times 10^{20}$ cm$^{-2}$ (K \kms)$^{-1}$ to
calculate the \h2 in CO--bright regions. The application of a constant
$X_{\mathrm {CO}}$ factor is only a rough estimate and only partly
successful. However, we found  no indications that the $f_c(T_{\mathrm
  {D}})$ conversion in this range could be invalid, thus this correction
appears to be valid in general for all diffuse \h2 regions.

Our $f_c(T_{\mathrm {D}})$ correction is applicable to an \h2 component
of the ISM that is termed CO--dark gas by \citep{Grenier2005}. This is
\h2 outside CO dominated regions.  For this diffuse molecular gas the
carbon is in the form of C or C$^+$ and not CO \citep{Wolfire2010}.  Our
approach allows us to quantify the atomic and molecular gas from the
diffuse atomic gas, via the diffuse molecular, up to the translucent
cloud regime \citep{Snow2006}. The $f_c(T_{\mathrm {D}})$ correction
according to Eq. \ref{eq:f_c} implies a moderate onset of \h2 formation
for \hi\ Doppler temperatures $ T_{\mathrm {D}} \la 1165$ K. In
consequence, the $f_c(T_{\mathrm {D}})$ term leads to a notable presence
of \h2 already in the canonical thin regions described in
Sect. \ref{Initial}, item 1.  Small-scale \hi\ structures are
filamentary and have a median Doppler temperature of $ T_{\mathrm {D}}
\sim 220$ K (\citet{Clark2014}, \citet{Kalberla2016}, and
\citet{Kalberla2018}). In this case $f_c(T_{\mathrm {D}}) = 2.55 $, thus
the fraction for the CO--dark \h2 interior to the CNM $f^N_{\mathrm
  {H2}} = 2~N_{\mathrm {H2}} / N_{\mathrm {H}} = 1 - f^{-1}_c =
0.61$. These CNM structures are clearly dominated by \h2. Filaments with
$ T_{\mathrm {D}} \la 220$ K cover 64\% of the sky. The bones of these
filaments, which we characterize according to Sect. \ref{bones} or
Fig. \ref{Fig_SkyMap_857} as $ T_{\mathrm {D}} \la 155$ K, have
$f^N_{\mathrm {H2}} \ga 0.68 $ and cover 48\% of the sky, thus in the
central parts of the filaments the CO--dark molecular gas is
enhanced. Only a few CNM clouds have $ T_{\mathrm {D}} \la 85$ K, but
these clouds are local condensations at prominent filamentary
structures. For $ T_{\mathrm {D}} \sim 50$ K about 90\% of the column
density is molecular, but only at 9\% of the sky positions. \hi\ Doppler
temperatures are upper limits to kinetic temperatures, hence clouds with
$ T_{\mathrm {D}} \la 85$ K are exceptionally cold in comparison to the
average \h2 rotational temperatures of 80 K   determined by
\citet{Savage1977} for diffuse \h2. We conclude that \hi\ filaments and
in particular their bones are cold, typically with $ T_{\mathrm{ex}}
\sim 50$ K, and associated with CO--dark molecular gas. Figure
\ref{Fig_Histo_vel} implies that most of the \h2 is encapsulated within
the \hi.

\begin{figure*}[thp] 
    \centering
    \includegraphics[width=18cm]{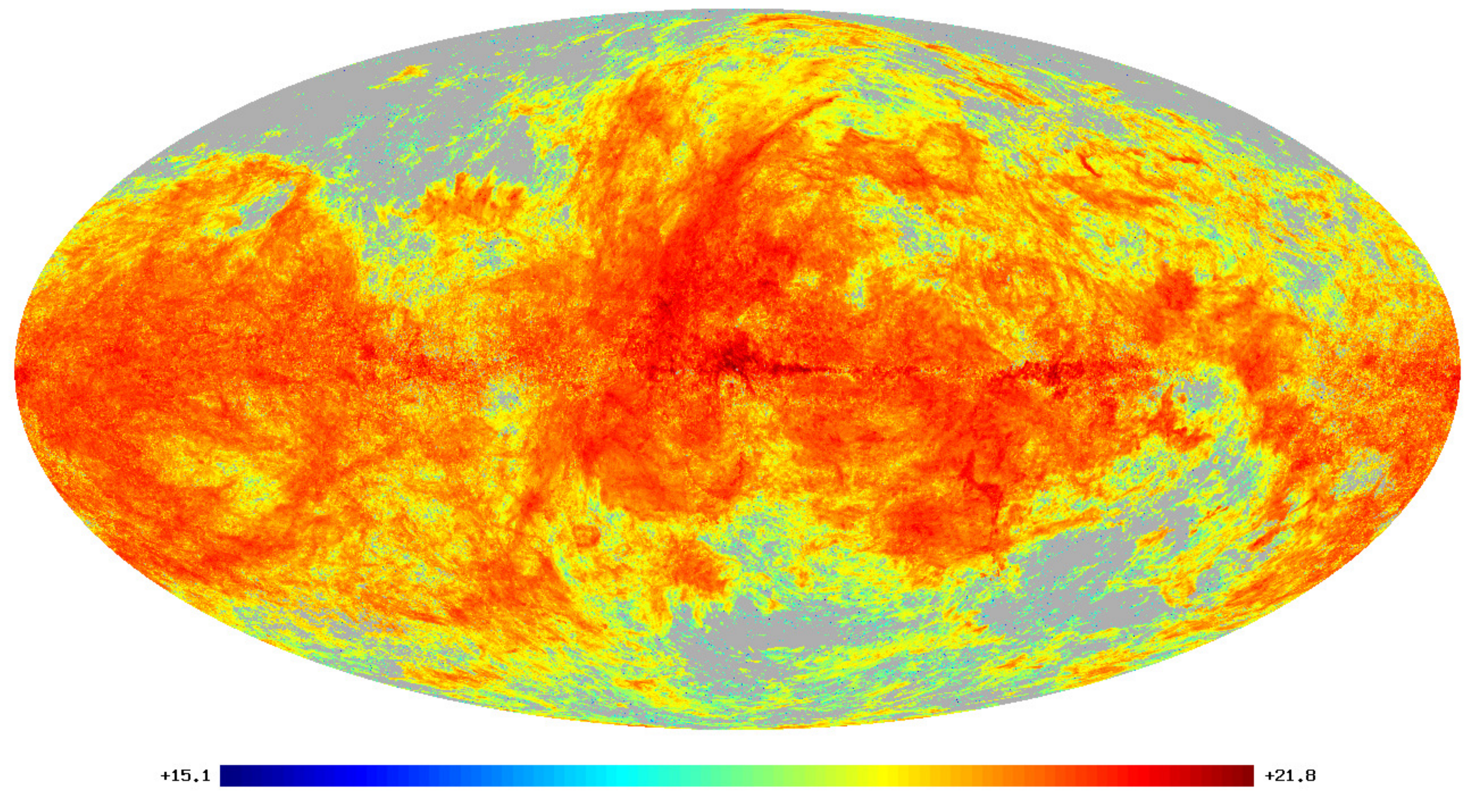}
    \caption{Distribution of CO--dark \h2 in the velocity range
      $|v_{\mathrm {LSR}}| \le 8$ \kms\ from the HI4PI
      survey. The scale is logarithmic, $N_{\mathrm {H}}$
      units are cm$^{-2}$. }
   \label{Fig_NH2_8}
\end{figure*}

\begin{figure*}[th] 
    \centering
    \includegraphics[width=18cm]{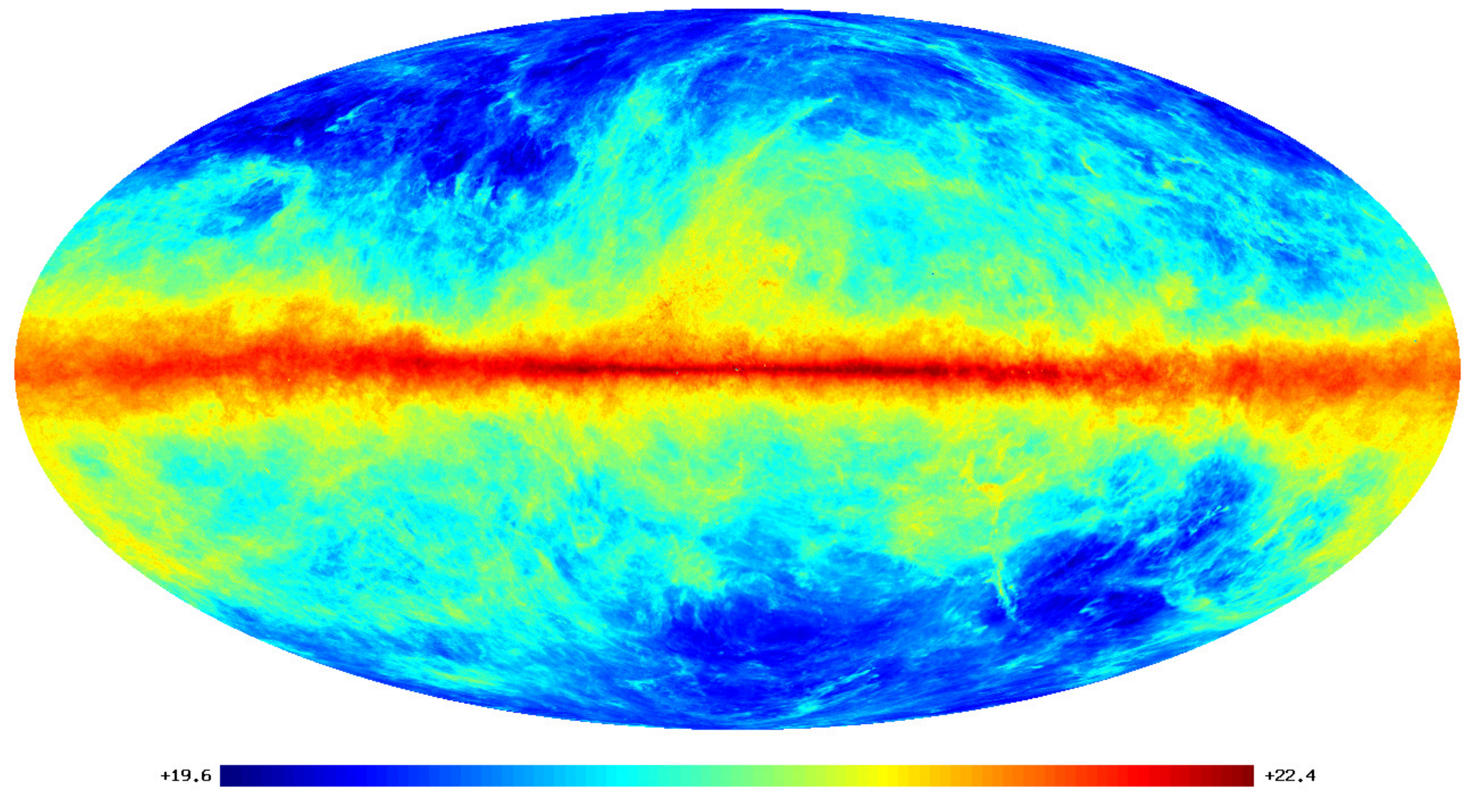}
    \caption{Distribution of \hi\ and diffuse CO--dark \h2 in the velocity range
      $|v_{\mathrm {LSR}}| \le 90$ \kms. The scale is logarithmic,
      $N_{\mathrm {H}}$ units are cm$^{-2}$. }
   \label{Fig_NH_90}
\end{figure*}

\begin{figure*}[th] 
    \centering
    \includegraphics[width=18cm]{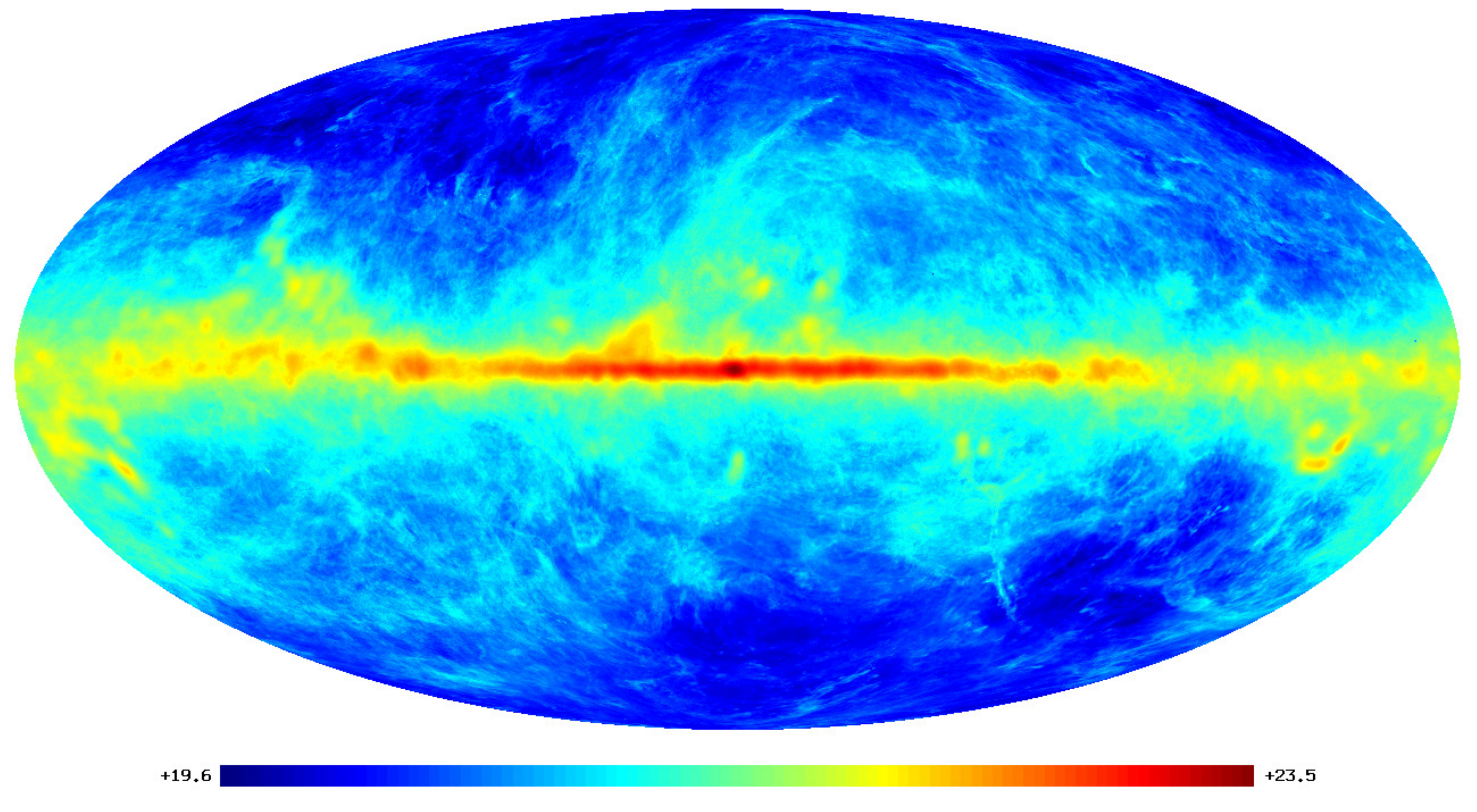}
    \caption{Distribution of \hi, CO--dark and estimated CO--bright \h2
      in the velocity range $|v_{\mathrm {LSR}}| \le 90$ \kms. The scale
      is logarithmic, $N_{\mathrm {H}}$ units are cm$^{-2}$. }
   \label{Fig_NH_CO_90}
\end{figure*}

Figure \ref{Fig_Plot_2D_hist} displays 2D histograms of the 
  frequency distribution for all Gaussian components with  CNM molecular
gas fractions $f^N_{\mathrm {H2}} = 2~N_{\mathrm {H2}} / N_{\mathrm {H}}
$ against the fractional $E(B-V)$ for each of these clouds.  We use 
  several  different selection criteria.  Already in the case of the
canonical thin regions with $N_{\mathrm {HI}} \le 4 ~ 10^{20}$ cm$^{-2}$
and $E(B-V) \le 0.08$ mag we find appreciable fractions $f^N_{\mathrm
  {H2}} \sim 0.5$, similar to the CO--masked gas, resulting in the
all-sky case in a slightly lower $f^N_{\mathrm {H2}} \sim 0.46$. The
implication is that CO--dark gas must be less abundant in the CO--masked
regions. Calculating the distribution there shows that the
$f^N_{\mathrm {H2}}$ ratio drops significantly for very obscured
regions. CO--dark \h2 gas is depleted in CO-rich regions since the
CO--dark part lies outside the dominant CO region
\citep{Wolfire2010}. Numerical simulations by \citet{Seifried2020MNRAS}
show that the dark gas fraction scales inversely with the amount of
well-shielded gas for $E(B-V) \ga 0.5$. This is the range where we find
a considerable drop in the dark gas fraction, on average $f^N_{\mathrm
  {H2}} \la 0.4$ (Fig. \ref{Fig_Plot_2D_hist}, bottom right).

Summing up all CNM components with $ T_{\mathrm {D}} \la 1165$ K we
obtain an all-sky  fraction of 46\% for the CO--dark \h2 gas;  excluding CO
dominated regions we get 49.6\%. Accounting for all gas in our local
vicinity, including the WNM, we find a molecular gas fraction of 18\%,
consistent with a fraction of 17\%, determined by
\citet{Savage1977}. This average molecular gas fraction compares to
fractions between 1\% and 30\% for cirrus clouds reported by
\citet{Gillmon2006}.  Our results are somewhat higher than the value of
27\% that can be inferred from \citet{Wolfire2010}.

All of the 2D histograms for the $N_{\mathrm {H}}/E(B-V)$ ratio in Fig.
\ref{Fig_Plot_regres} show a bending of the slope of the distributions
at a column density $5 ~ 10^{20} \mathrm {cm}^{-2}$. This effect was
first observed by \citet{Savage1977} and attributed to a systematic
change in the molecular gas fraction at this column density. We find
that the onset of \h2 formation is not limited to a threshold in column
density. Only the Doppler temperature of the gas is important, and
Fig. \ref{Fig_Plot_2D_hist} (top left) is explained by the fact that a
significant fraction of this gas does not reach low Doppler
temperatures.

Figure \ref{Fig_Plot_2D_hist} indicates that the fraction $f^N_{\mathrm
  {H2}}$ must be relatively constant for the bulk of the CNM
components. The investigations by \citet{Wolfire2010} appear to be
consistent with this finding. They use a different definition for the
molecular gas fraction, but their conclusion is that the fraction of
molecular mass in the dark component is remarkably constant and
insensitive to the incident ultraviolet radiation field strength and the
internal density distribution, and the mass of the molecular cloud. Our
empirical $f_c(T_{\mathrm {D}})$ correction is based on a statistical
investigation over a large number of positions. We can only  claim that
the $f_c(T_{\mathrm {D}})$ correction is valid on average. If however
the molecular gas fraction is as insensitive to environmental conditions
as claimed by \citet{Wolfire2010}, the correction can faithfully be
applied even to individual CNM clouds.

We display in Fig. \ref{Fig_NH2_8} a map of the CO--dark gas for
$|v_{\mathrm {LSR}}| \le 8$ \kms. This velocity range is most closely
representative of filamentary structures in our local vicinity
\citep[][Sect. 5.13]{Kalberla2016}. We see that most of the CO--dark
molecular gas is organized in filaments. For a comparison with the total
amount of \hi\ and CO--dark \h2 in the velocity range $|v_{\mathrm
  {LSR}}| \le 90$ \kms\ see   Fig. \ref{Fig_NH_90}. Except for
CO--bright regions, which  were disregarded here because of the uncertain
$X_{\mathrm {CO}}$ correction, this map may be helpful to supplement
interstellar reddening maps. Figure \ref{Fig_NH_CO_90} displays our
estimate of the total \hi, CO--dark and CO--bright gas distribution in the
Milky Way. Figures \ref{Fig_NH2_8} to \ref{Fig_NH_CO_90} 
demonstrate our current estimates on the distribution of diffuse \h2 and
the total neutral hydrogen in the Milky Way, without and with CO--bright
\h2. The significance of these results needs to be evaluated.

\subsection{\h2 power spectra }
\label{H2_power}

\begin{figure}[thp] 
    \centering
 \includegraphics[width=8.5cm]{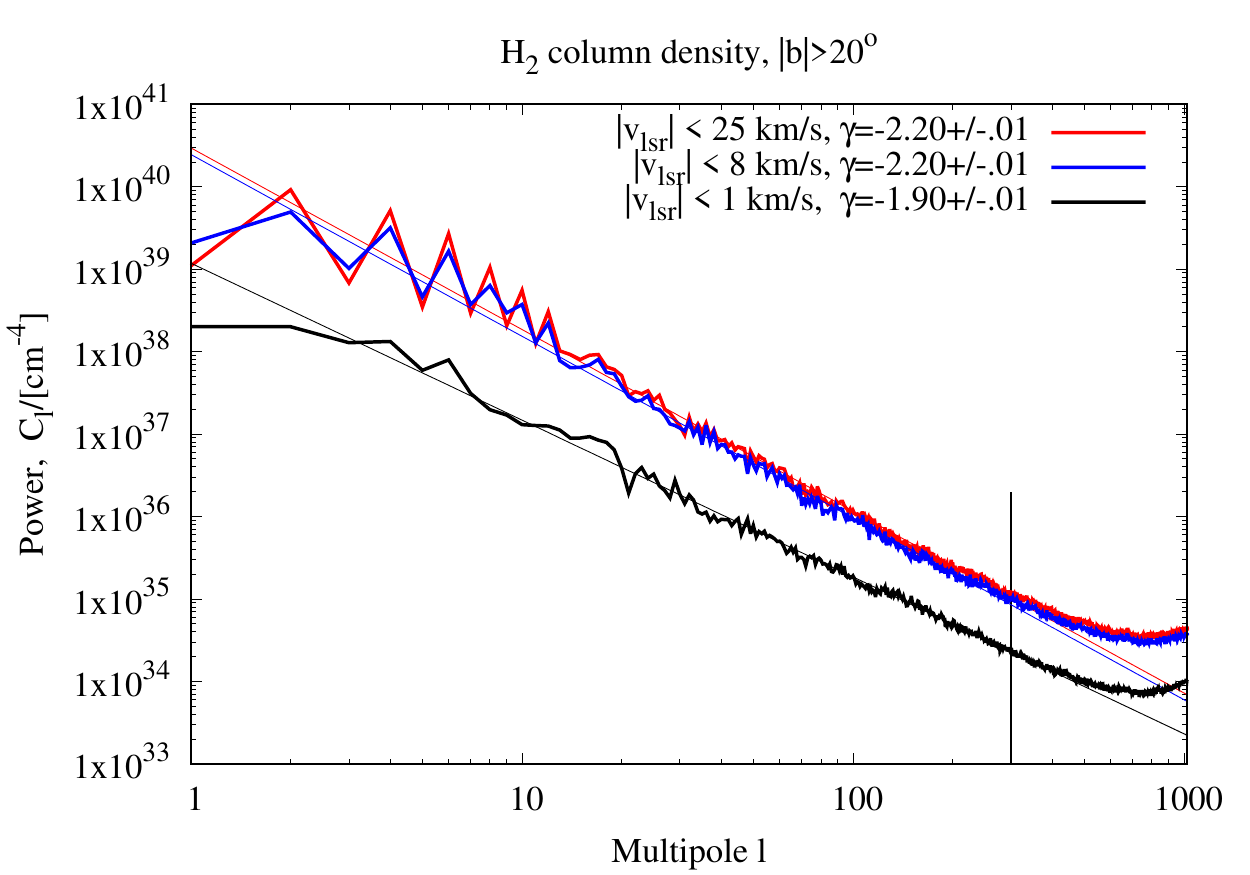}
 \includegraphics[width=8.5cm]{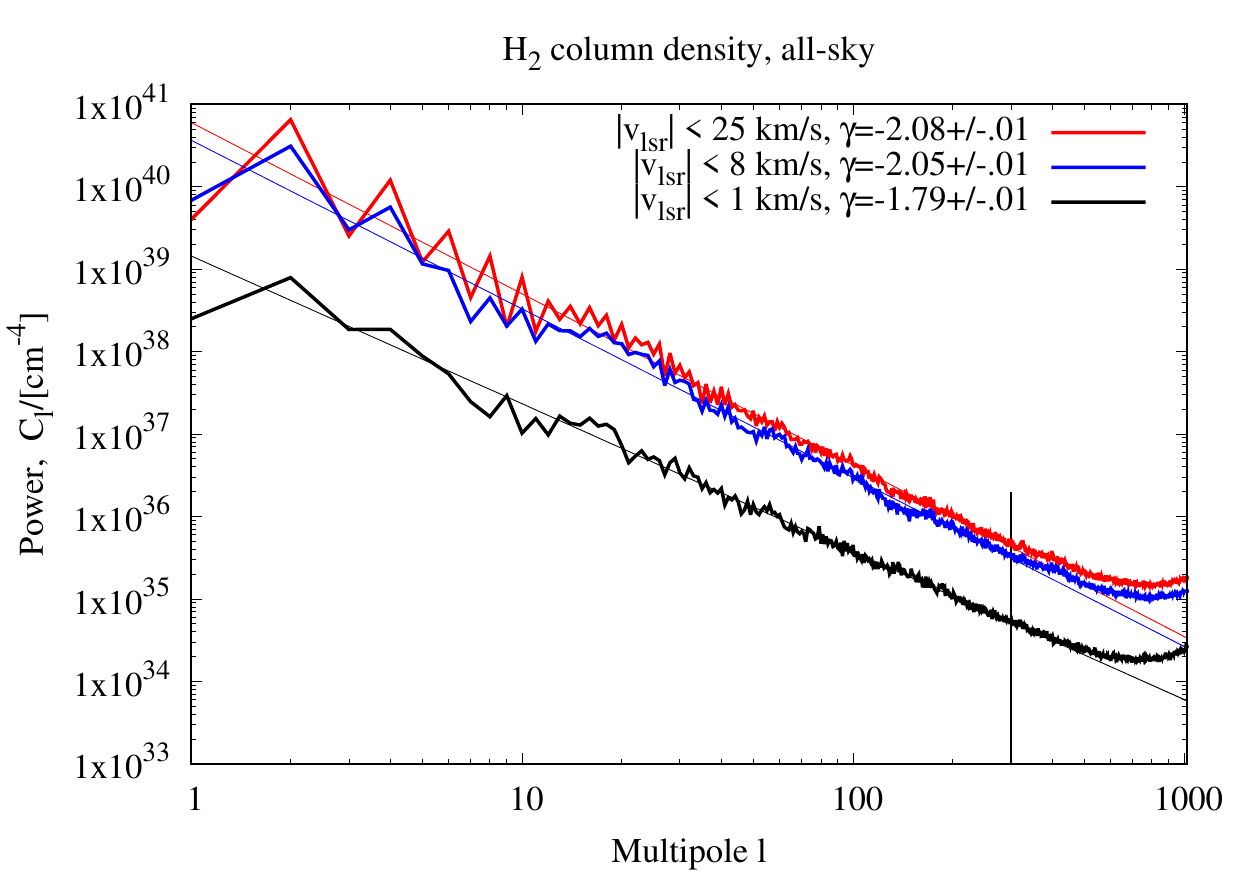}
    \caption{Power spectra for the derived column density distributions
      of the diffuse \h2 at $|v_{\mathrm {LSR}}| < $ 25, 8, and 1 \kms\
      at high latitudes (top) and all-sky (bottom). The power indices
      $\gamma$ were fitted for multipoles $ 10 < l < 300$; the vertical
      line indicates the upper limit in $l$. }  
   \label{Fig_Plot_power}
\end{figure}

The \h2 distribution displayed in Fig. \ref{Fig_NH2_8} is linked to the
\hi\ distribution via Eq. \ref{eq:f_c}. This implies that the diffuse
\h2 is embedded in the \hi. The \hi\ filaments that host the \h2 must 
necessarily be density structures in conflict with the interpretation of
filaments as velocity caustics (\citet{Lazarian2000},
\citet{Lazarian2018}, and \citet{Yuen2019}). We can use HI4PI data to
estimate the power distribution of the diffuse molecular \h2 at large
scales. Our data processing is the same as described by
\citet{Kalberla2019} except that we restrict our analysis to the diffuse
CO--dark \h2.  Figure \ref{Fig_Plot_power} shows the power spectra for
three velocity windows: $|v_{\mathrm {LSR}}| < $ 25, 8, and 1 \kms. On
top we display spectra for $|b| > 20\deg$, on the bottom are all-sky
data. These spectra are shallow in comparison to the CNM power spectra
shown in Figs. 1 and 23 of \citet{Kalberla2019}. The diffuse \h2 is
embedded in the CNM and the colder the \hi, the more pronounced are
intermittent small-scale structures with transitions to \h2.  The
relations between hierarchical scaling of successive structures in a
turbulent medium have been described by \citet{She1994}. A
3D incompressible flow is considered as a hierarchy of
structures, the most singular structures are assumed to be
filaments. Accordingly high intensity structures are understood as
filaments with a Hausdorff dimension of one. Reducing the dimension
implies a reduction of the power law index. The expected energy spectrum
of turbulence is accordingly $E(k) \propto k^{-5/3 -0.03}$.

Figure \ref{Fig_Plot_power} indicates that most of the \h2 structures
are local; $|v_{\mathrm {LSR}}| < 8 $ \kms\ contains a significant
fraction of the power. \h2 structures extracted for $|v_{\mathrm {LSR}}|
< 1$ \kms\ are the best examples with the highest signal-to-noise ratio for
an all-sky distribution of intermittent filamentary structures. The
corresponding power spectra with fitted power law indices $-1.79 \ga
\gamma \ga -1.9$ are significantly shallower than $ -2.14 \ga \gamma \ga
-2.37 $ for the CNM \citet[][Fig. 1]{Kalberla2019}. These results
support the conjecture by \citet{She1994}, who find that the  nature of these
asymptotic flow structures is a specific property of the
three-dimensional incompressible flows and that only filamentary structures seem
to be mechanically stable. In the case of the diffuse ISM, phase
transitions increase intermittency,  therefore affecting the properties
of the turbulent flow. Spectral indices for the \hi\ distribution in
narrow velocity channels were found to depend on Doppler temperatures,
and \citet{Kalberla2016,Kalberla2017} and \citet{Kalberla2020}
interpret this as an indication that the turbulent flow is affected by
phase transitions. The coupling of linear density structures to the
Galactic magnetic field \citep{Clark2019b} appears in addition to
support the hierarchy of structures in the ISM that approach Hausdorff
dimensions of one for filamentary structures.

\begin{figure}[thp] 
    \centering
 \includegraphics[width=8.5cm]{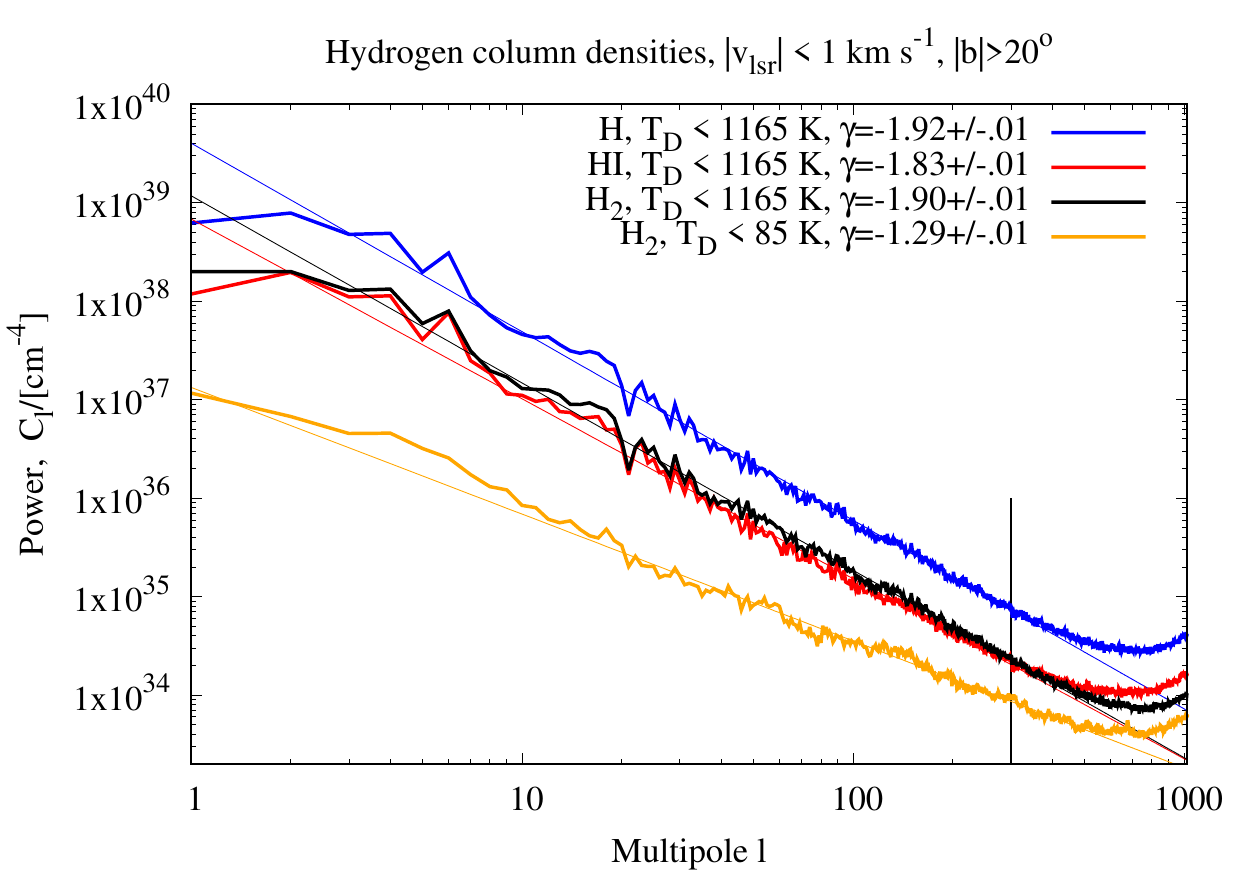}
 \includegraphics[width=8.5cm]{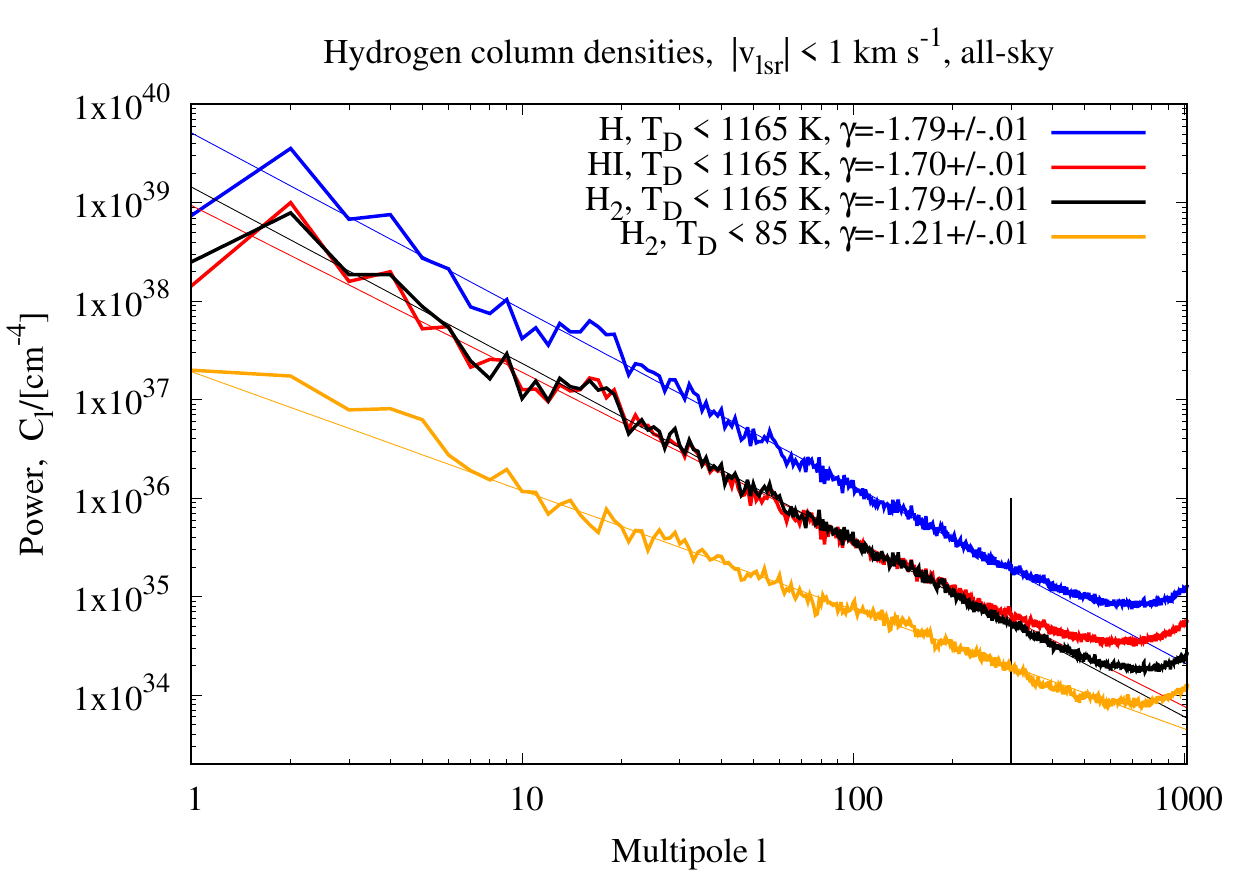}
    \caption{ Power spectra, comparing column density distributions for
      \hi, H, and \h2 (CO--dark) at $|v_{\mathrm {LSR}}| < 1 $ \kms\ for
      Doppler temperatures $T_{\mathrm {D}} < 1165$ K and $T_{\mathrm
        {D}} < 85$ K at high latitudes (top) and all-sky (bottom). The
      power indices $\gamma$ were fitted for multipoles $ 10 < l < 300$;
     the vertical line indicates the upper limit in $l$. } 
   \label{Fig_Plot_power_2}
\end{figure}

Figure \ref{Fig_Plot_power_2} compares power spectra selecting different
neutral hydrogen species, \hi, H, and \h2 (CO--dark gas only) from our
model calculations. As in Fig. \ref{Fig_Plot_power} we selected
components with Doppler temperatures $T_{\mathrm {D}} < 1165$ K at
$|v_{\mathrm {LSR}}| < 1 $ \kms. We compare power spectra at high
latitudes (top) and all-sky (bottom). The column density spectra for $
N_{\mathrm {H}} = N_{\mathrm {HI}} + 2 ~ N_{\mathrm {H2}}$ (blue) have the
highest power,
as expected since this is the total amount of all neutral hydrogen. The $(2 ~ N_{\mathrm {H2}})$ power spectra (black) have
within the uncertainties power law slopes that are identical to the $
N_{\mathrm {H}} $ power spectra and are straight up to high multipoles
$l \sim 400$. The high power tails of the \hi\ spectra (red) bend up at
$l \ga 300$. Comparing these \hi\ spectra with
\citet[][Fig. 1]{Kalberla2019} indicates systematic differences. The
turn-over of these CNM power spectra with deviations from a fitted
straight power law is ever earlier at $l \ga 100$. Furthermore, those CNM
power spectra are significantly steeper ($-2.14 \ga \gamma \ga -2.37 $)
than the \hi\ spectra from Fig. \ref{Fig_Plot_power_2} constrained by
$T_{\mathrm {D}} < 1165$ K ($-1.70 \ga \gamma \ga -1.83$).

In addition to power spectra with Doppler temperatures $T_{\mathrm
    {D}} < 1165$ K, we plot in Fig. \ref{Fig_Plot_power_2} $(2 ~
  N_{\mathrm {H2}})$ power spectra for $T_{\mathrm {D}} < 85$ K
  (orange); this is for the steep branch from Eq. \ref{eq:f_c} and
  Fig. \ref{Fig_Tss_fit}. Even though the \h2 clumps from the
  coldest CNM are located along \h2 filaments, most of the filamentary
  structures are broken and this sample of \h2 clumps approaches a
  random distribution for objects with a Hausdorff dimension of zero. In
  agreement with the conjecture by \citet{She1994}, these power spectra are
  very shallow. These power spectra are straight up to $ l \la 500$.
  Deviations, caused by optical depth effects, are not recognizable.

For an interpretation of all these different power spectra we need to
take into account that different ISM phases (WNM, LNM, CNM, and \h2) are
related to each other. A complete description of these phases demands
that all cross-correlations between the phases need to be taken into
account \citep[][Eq. 4]{Kalberla2019}. A detailed treatment of the cross-correlations is beyond the scope of the current publication, but we may
safely conclude that the $(2 ~ N_{\mathrm {H2}})$ power spectra (black)
in Fig. \ref{Fig_Plot_power_2} have the best-defined self-similar
straight power spectra for the cold ISM in the intermediate range of
scales up to $l \la 400$. We argued in Sect. \ref{Fitting_fc} that the
definition for Gaussian CNM components \citep{Kalberla2019} is broadly
consistent with an upper limit of Doppler temperatures $T_{\mathrm{D}}
\la 1165$ K. However, from the outstanding properties of the $(2 ~
N_{\mathrm {H2}})$ spectral power distributions from
Fig. \ref{Fig_Plot_power_2} it appears appropriate to consider this
limit as relevant for phase transitions that lead to significant
filamentary \h2 structures.  This threshold should not be considered as
a canonical value; instead, it is  a weighted mean comprising a best fit
value for all physical conditions considered by us over 74\% of the sky
outside CO--bright regions.

A comparable upper limit $T_{\mathrm{D}} \sim 1100$ K was found for
Doppler temperatures of filamentary features derived by unsharp masking
of HI4PI data by \citet[][Sect. 5.11]{Kalberla2016}. At a Mach number of
3.7 it is consistent with an upper limit of $T_{\mathrm {kin}} \sim 220$ K
for the kinetic temperature as expected for a stable CNM phase
\citep{Wolfire2003}. An upper $T_{\mathrm {D}}$ limit for
$f_c(T_{\mathrm {D}})$ is not in conflict with phase transitions out of
the WNM but it implies that the \hi\ must first cool down to the CNM
before \h2 molecules can form.

\section{Summary and conclusion}
\label{Summary}

We use Doppler temperatures $T_{\mathrm {D}}$ from a Gaussian
decomposition of HI4PI data to study temperature dependences of
$E(B-V)/N_{\mathrm {HI}}$. This ratio increases with $-\log(T_{\mathrm
  {D}}),$ and we interpret this trend with the presence of unaccounted
molecular hydrogen: $E(B-V)/N_{\mathrm {HI}} \ga E(B-V)/(N_{\mathrm
  {HI}}+ 2~N_{\mathrm {H2}})$. Systematic changes of
$E(B-V)/N_{\mathrm {HI}}$ allow the definition of a temperature
dependent correction under the assumption of a constant
$E(B-V)/(N_{\mathrm {HI}}+2 ~ N_{\mathrm {H2}})$ ratio.

This empirical $f_c(T_{\mathrm {D}})$ correction (Eq. \ref{eq:f_c})
allows us to estimate the dark molecular gas content in the CNM. We find
evidence of \h2 at temperatures $T_{\mathrm {D}} \la 1165$ K; on
average the diffuse molecular gas fraction in CNM clouds outside
CO--bright regions is $f^N_{H2} = 2 ~ N_{\mathrm {H2}} / N_{\mathrm {H}}
= 0.46$.  Filamentary \hi\ structures with $T_{\mathrm {D}} \la 220$
  K are cold with $ T_{\mathrm{ex}} \la 50$ K, are with $f^N_{H2} \ga
  0.61 $ dominated by \h2, and are associated with dust. According to
  \citet{Clark2019b} they are aligned with the magnetic field,
  representing magnetically coherent regions of space. All these
  filaments have column densities below the limit $ \log ( N_{\mathrm
    {H}} / {\mathrm {cm^{-2}}}) \sim 21.7 $ where the preferential
  orientation of the magnetic field along the filaments switches to
  being perpendicular to the $N_{\mathrm {H}}$ contours
  \citep{Planck2016b}. We find evidence that the central parts of the
  filaments, the bones, have the lowest temperatures and increased
  molecular gas fractions $f^N_{H2}$.  The \hi\ encapsulates 
  the \h2. Applying the $f_c(T_{\mathrm {D}})$ correction leads to a
  significant reduction of systematic deviations in the
  $E(B-V)/(N_{\mathrm {HI}}+2 ~ N_{\mathrm {H2}})$ ratio at high
  Galactic latitudes.  Extending the $T_{\mathrm {D}}$ correction to
CO--bright regions is possible with some limited success for a factor
$X_{\mathrm {CO}} = 4.0 \times 10^{20}$ cm$^{-2}$ (K \kms)$^{-1}$.

Our empirical correction is based on statistical investigations,
covering 74\% of the sky; thus, the validity is limited to an on-average
correction. However theoretical investigations by \citep{Wolfire2010}
indicate that the \h2 content is insensitive to environmental
conditions. If this is really the case, our corrections may be universal
and can be used to predict the foreground attenuation affecting our view
to the distant universe. It appears worth  trying this because, as they say,  the proof of
the pudding is in the eating, and  we provide the necessary data\footnote[7]{\url{https://www.astro.uni-bonn.de/hisurvey/}}.

\begin{acknowledgements}
  We thank the referee for careful reading and constructive criticism.
  U. H. acknowledges the support by the Estonian Research Council grant
  IUT26-2, and by the European Regional Development Fund (TK133).  HI4PI
  is based on observations with the 100-m telescope of the MPIfR
  (Max-Planck-Institut für Radioastronomie) at Effelsberg and the Parkes
  Radio Telescope, which is part of the Australia Telescope and is
  funded by the Commonwealth of Australia for operation as a National
  Facility managed by CSIRO. This research has made use of NASA's
  Astrophysics Data System. Some of the results in this paper have been
  derived using the HEALPix package. We also used the Karma package by
  R.E. Gooch.
\end{acknowledgements}

\end{document}